\documentclass[prd,nofootinbib,preprintnumbers,preprint]{revtex4}

\usepackage{indentfirst}

\usepackage{bm}
\usepackage{mathrsfs}
\usepackage{comment}
\usepackage{amsfonts}
\usepackage{amssymb}
\usepackage{dsfont}
\usepackage{euscript}
\usepackage{graphicx}
\usepackage[hyperfootnotes=false]{hyperref}
\usepackage[dvipsnames]{xcolor}
\usepackage{tensor}
\usepackage{amsmath,latexsym}
\usepackage{cmap}

\def\be{\begin{equation}}
\def\ee{\end{equation}}
\def\ba{\begin{eqnarray}}
\def\ea{\end{eqnarray}}

\begin{document}
\title{ 
Gauged supergravities: solutions with Killing tensor}
\author{Dmitri Gal'tsov$^{a}$  \email{galtsov@phys.msu.ru}}
\author{Rostom Karsanov$^{a}$\email{karsanovrz@my.msu.ru}}
\affiliation{$^{a}$Faculty of Physics, Moscow State University, 119899, Moscow, Russia}

\begin{abstract}
We perform full integration of the stationary axisymmetric Einstein-Maxwell-dilaton-axion (EMDA) theory with and without potential using a recently proposed generalization of Carter's approach to spacetimes beyond type D, allowing the Killing tensor. Crucial to our construction is a new parametrization of the dilaton and axion fields based on the analyticity argument. The general solution in the ungauged case is asymptotically locally flat and contains two more parameters compared to EMDA black holes previously obtained using Harrison transformations. In the gauged case, the general solution is asymptotically AdS and includes flat and hyperbolic topological solutions, as well as generalization of the Kerr-Sen-AdS metric  with three additional parameters. Our approach can be applied to more general four-dimensional ungauged and gauged supergravities.

\end{abstract}
 \maketitle
\setcounter{page}{2}

\setcounter{equation}{0}
\setcounter{subsection}{0}
\setcounter{section}{0}
\setcounter{equation}{0}
\setcounter{subsection}{0}
\setcounter{section}{0}

\section{Introduction}\label{intro}

\setcounter{equation}{0}
There is a widespread interest to constructing rotating black hole solutions in extended four-dimensional supergravities with vector and scalar fields. While supersymmetric solutions can be constructed using Bogomolnyi equations, for more general solutions special techniques are required. An efficient method consists in using dimensional reduction of the supergravity actions on stationary spacetimes to three-dimensional sigme-models.   Originating in vacuum \cite{Ernst:1967wx} and electrovacuum \cite{Ernst:1967by,Neugebauer:1969wr} gravity, in the Kaluza-Klein theory \cite{ Neugebauer,Maison:1979kx,Clement:1986kzn,Breitenlohner:1987dg}, it was extended to the Einstein-Maxwell-dilaton-axion theory (EMDA, the main subject here) in \cite{Galtsov:1994pd,Galtsov:1994sjr}. A general treatment of supergravity sigma models was given in \cite{Breitenlohner:1998cv}, applications to specific four-dimensional supergravities were considered in \cite{Galtsov:1994pd,Youm:1997hw, Bouchareb:2007ax,Chow:2013tia, Chow:2014cca,Bogush:2020obx} (most notably in \cite{Chow:2014cca}) and many other papers. The symmetries of the sigma models included Harrison transformations that allowed charged supergravity black holes to be generated from the Kerr metric. 

However, this approach fails in gauged supergravities, where a scalar potential is present that destroys the sigma model structure of the three-dimensional theory. While spherically symmetric solutions can be easily obtained, rotating ones have been constructed only by guesswork inspired by known electrovacuum analogues, followed by computer verification \cite{Chong:2004na,Chow:2010sf,Chow:2010fw,Klemm:2012vm,Chow:2013gba,Gnecchi:2013mja,Wu:2020cgf,Gallerati:2021cty,Anabalon:2024cnb,Zhu:2024jhw}. 
A notable exception is the pure ${\cal N}=2$ theory without vector multiplets, which is in fact an Einstein-Maxwell theory with a cosmological constant. Such solutions were obtained analytically by Carter in 1968 \cite{Carter:1968ks} and later independently by Plebanski \cite{Plebanski:1975xfb} using a different technique that can be applied to spacetimes admitting a second-rank Killing tensor in addition to stationarity and axial symmetry. It was shown that such solutions of the Einstein-Maxwell equations with a cosmological constant can be found analytically, without resorting to the technique of sigma models at all. These solution belonged to Petrov type D. Later, Carter's approach was widely applied to four-dimensional and multi-dimensional spacetimes admitting generalized Killing-Yano structures \cite{Frolov:2017kze}. 

Recently \cite{Galtsov:2024vqo}, the Carter parametrization of the metric was rederived from the Benenti-Francaviglia (BF) ansatz \cite{Benenti:1979erw} for metrics admitting the Killing tensor by imposing certain restrictions on BF functions (see also \cite{Anabalon:2016hxg}). These restrictions guarantee the existence of two null geodesic shearfree congruences and the separability of the Klein-Gordon equation without the assumption that the metrics are algebraically special. The resulting metrics are in general of type $I$, or more precisely, of a type $I_{B}$ sector whose algebraically special subspace consists of metrics of type D only. It was shown that all known black holes of ungauged four-dimensional supergravities belong to this class.
Moreover, it turns out that known solutions to gauged supergravities also possess a Killing tensor,
though they do not belong to Petrov type D. 

The aim of this paper is to perform a full integration of a truncated version of ${\cal N}=4$ supergravity (EMDA theory) and to extend the approach to the theory with the potential arising in gauged ${\cal N}=4$ supergravity. The presence of  dilaton and  axion makes the theory more complicated than the Einstein-Maxwell theory, and its successful integration is made possible by a suitable parameterization of the complex axidilaton field consistent with the metric ansatz. Such a parameterization was not known so far, and we derive it here from an analyticity argument. 

The basic strategy, borrowed from Carter, is to extract from the entire system some linear equations that reveal the polynomial structure of the BF coefficient functions, and then solve the remaining nonlinear coupled equation by fixing the relations between the polynomial coefficients. We apply this to both ungauged and gauged EMDA-s.
In the ungauged theory we derive a general solution that extends the seven-parameter family of EMDA black holes \cite{Galtsov:1994pd} found earlier using the sigma model technique. In the gauged case, we obtain a solution that confirms and generalizes the Kerr-Sen-AdS black hole \cite{Wu:2020mby,Sakti:2022izj,Sakti:2022txd,Ali:2023ppg}, endowing it with new parameters and presenting for the first time a complete analytical derivation. Our solutions include topological ones that should appear in the AdS asymptotic case   \cite{,Lemos:1994xp,Cai:1996eg,Birmingham:1998nr,Vanzo:1997gw,Brill:1997mf,Klemm:1997ea,Gnecchi:2013mja,Zhu:2024jhw}. 

The outline of the paper is as follows. In Section 2 we present the geometric structure for stationary axisymmetric spacetimes admitting a second-rank Killing tensor and compute the corresponding Ricci tensor in tetrad form. In Section 3 the action and equations of motion of the gauged EMDA are discussed in real and complex terms. Section 4 is devoted to extracting linear equations for the BF coefficient functions, from which their polynomial structure becomes apparent. In Sections 5 and 6 the final integration of the ungauged and gauged EMDA is performed and the properties of the solutions obtained are briefly discussed. Our results are summarized in the Conclusion.

\section{Integrable spacetime}        
\subsection{Benenti-Francavigila ansatz}
The class of metrics introduced in \cite{Galtsov:2024vqo} and called $I_B$ (Benenti) is defined as follows.
Starting with stationary axisymmetric metric is written in the coordinates $x^\mu=(x^a,\,x^i)$, where $x^a=t,\, \varphi$ correspond to the subspace spanned by the Killing vectors $K^{(t)}=\partial_t$  and  $K^{(\varphi)}=\partial_\varphi$ and $x^i=r,\,y$, belong to orthogonal two-dimensional space whose metric without loss of generality can be assumed diagonal. 
The contravariant metric is parameterized as
$g^{\mu\nu}=\left(g^{ab},\;g^{ij}\right)$ : 
\begin{equation}\label{MetrUpBeneti}
g^{ab}=\Sigma^{-1}
\begin{pmatrix}
  A_3-B_3 \; & \; A_4-B_4\\   A_4-B_4\; & \;A_5-B_5
  \end{pmatrix}
,\qquad g^{ij}=-\Sigma^{-1}\begin{pmatrix}
  A_2&0\\  0& B_2
  \end{pmatrix}  
\end{equation}
where two sets of arbitrary functions are introduced $A_k(r),\,B_k(y),$ $k=1..5$ depending each on one variable, $r$ and $y$ respectively. In order to ensure existence of an exact Killing tensor, the conformal factor  $\Sigma=\Sigma(r,y)$ must be separable
$\Sigma=A_1(r)+B_1(y)$.
Then the Killing tensor, satisfying the equation
$ \nabla_{(\alpha}K_{\mu\nu)}=0,$ can be written in ``slice-reducible''
 form \cite{Kobialko:2022ozq, Kobialko:2021aqg}:
\begin{equation}\label{eq:K_r}
   K^{\mu\nu}=
   -A_1 g^{\mu\nu}
   -A_2\delta^\mu_r \delta^\nu_r
   +\tilde K^{\mu\nu}_r,
\end{equation}
where
\begin{equation}\label{KKV1}
\tilde K^{\mu\nu}_r=A_3\delta^\mu_t \delta^\nu_t +2A_4\delta^{(\mu}_t \delta^{\nu)}_\varphi+A_5\delta^\mu_\varphi \delta^\nu_\varphi.  
\end{equation}
The first term in (\ref{eq:K_r})  is trivial Killing tensor on the hypersurfaces ${\cal S}_r$ corresponding to $r={\rm const}$. The second term is orthogonal to ${\cal S}_r$ and thus irrelevant, while the third term $\tilde K^{\mu\nu}_r$ is a reducible Killing tensor on this hypersurface, representable as a linear combination of the tensor products of the Killing vectors projected onto it.  Similarly, this Killing tensor can be presented as slice-reducible with respect to foliation by the hypersurafaces of constant $y$. For an arbitrary conformal factor $\Sigma$ only a conformal Killing tensor exists.

It is
assumed that in a significant region of space-time (e.g. beyond the horizon or the ergosphere) all the BF coefficient functions are positive, and  analytic continuation into the negative region can be done in the usual way.  Other sign conditions follow from the metric signature: with the same reservations 
    $A_3-B_3>0,\; A_5-B_5<0,\; \Sigma>0$.
 
The static limit of the Benenti-Francaviglia ansatz corresponds to 
\begin{equation}
    A_4\equiv 0,\quad B_4\equiv 0.
\end{equation}
\subsection{Constraints}
The gauge freedom inside the BF ansatz consists of two coordinate transformations 
   $r\rightarrow \tilde{r}(r),\;y\rightarrow \tilde{y}(y),$
 containing two arbitrary functions of independent variables. These can be used to impose additional conditions on $A_2,\; B_2$:
\begin{equation}\label{ab}
A_2  A_5 =  a^2={\rm  const},\quad   B_2  B_5 =  b^2={\rm  const}.
\end{equation}
In what follows we will assume the validity of this gauge.  

Next, we impose two constraints 
on the Benenti coefficients  \cite{Galtsov:2024vqo}:
  \begin{align}
\label{435}
A_4=\sqrt{A_3 A_5} ,\qquad B_4=\sqrt{B_3 B_5},
\end{align}
which guarantee the existence of two null shearfree geodesic congruences. For vacuum and electrovacuum, this property together with existence of the Killing tensor would mean (Goldberg-Sachs theorem \cite{goldberg1962theorem}) that the metric will be of Petrov type D. With more general matter sources, one has to deal with metrics of type I, more precisely, with the subclass $I_B$, whose algebraically special subspace consists only of type D. As was shown in \cite{Galtsov:2024vqo}, BF ansatz with constraints (\ref{435}) is suitable for this.

In the static limit the relations (\ref{435}) degenerate. A closer look shows that  a correct choice of  static limit will be
\begin{equation}
    A_5\equiv 0,\quad B_3\equiv 0,\quad A_3\neq 0,\quad B_5\neq 0,
\end{equation}
implying $a=0, \,b\neq 0.$

One more constraint on BF functions is intended to ensure separability of the Klein-Gordon equation. It reads  $\sqrt{-g}=\Sigma$ \cite{Galtsov:2024vqo}.
This  can be rewritten in the gauge (\ref{ab}) as
\begin{equation}\label{sigConsAlt}   \Sigma= A_1+B_1= bA_{23}-aB_{23},
\end{equation}
where \begin{equation}
A_{23}=\sqrt{A_2A_3},\quad\quad B_{23}=\sqrt{B_2B_3}.
\end{equation}  Since all $A$ depend only on $r$, and all $B$ depend only on $y$, it is easy to see that the derivatives with respect to the corresponding arguments are related as follows:
\begin{equation}
   \label{A1B1} A_1'=bA_{23}',\qquad B_1'=-aB_{23}'.
\end{equation}

With all the above constraints imposed, one obtains the following metric parametrization:
\begin{equation}\label{metricmain0}
    ds^2=\frac{A_2}{\Sigma}(b dt - B_{23}d\varphi)^2 - \frac{B_2}{\Sigma}(adt-A_{23}d\varphi)^2-\frac{\Sigma}{A_2}dr^2-\frac{\Sigma}{B_2}dy^2,
\end{equation}
This coincides with Carter's ansatz \cite{Carter:1968ks} which allowed  to solve vacuum and electrovacuum Einstein equations, obtaining general solutions admitting a Killing tensor. Here we apply this ansatz together with appropriate parameterizations of the vector field, dilaton and axion to both ungauged and gauged EMDA theory with a potential. 
\subsection{Tetrad formalism}
For our purposes, it will be more convenient to proceed with calculations using tetrads
$ 
g_{\mu\nu} =\eta_{ab}e_\mu^a e_\nu^b,
$ 
where $\eta_{ab}={\rm diag}(1,-1,-1,-1)$ for $a,b=1,2,3,4$ and the natural one-form basis is
\begin{equation}\label{tetrad}
    \begin{aligned}
        e^1&=\alpha(bdt-B_{23}d\varphi),\\
        e^2&=\beta(adt-A_{23}d\varphi),\\
        e^3&=\alpha^{-1}dr,\\
        e^4&=\beta^{-1}dy,
    \end{aligned}
\end{equation}
with  
\begin{equation}
\alpha=\sqrt{A_2/\Sigma}, \qquad\beta=\sqrt{B_2/\Sigma}.
\end{equation}
Note that one of the constants $a,b$ can be fixed by time rescaling. Since $a=0$ corresponds to the static case, this parameter is worth to be kept arbitrary. The constant $b$ can be set equal to one without loss of generality, but we will leave it in order to keep the $A-B$ symmetry explicit until the end of the calculations.

Tetrad components of the Ricci tensor read \cite{Landau:1988}:
\begin{equation}\label{RicciLL}
\begin{aligned}
    R_{ab}&=-\frac{1}{2}\Big(\tensor{\lambda}{_{ab}^c_{,c}}+\tensor{\lambda}{_{ba}^c_{,c}}+\tensor{\lambda}{^c_{ca}_{,b}}+\tensor{\lambda}{^c_{cb}_{,a}}+\tensor{\lambda}{^{cd}_{b}}\tensor{\lambda}{_{cda}}+\tensor{\lambda}{^{cd}_{b}}\tensor{\lambda}{_{dca}}-\\
    &-\frac{1}{2}\tensor{\lambda}{_{b}^{cd}}\tensor{\lambda}{_{acd}}+\tensor{\lambda}{^{c}_{cd}}\tensor{\lambda}{_{ab}^d}+\tensor{\lambda}{^{c}_{cd}}\tensor{\lambda}{_{ba}^d}\Big),
    \end{aligned}
\end{equation}
where the quantities $\lambda_{abc}=-\lambda_{acb}$ are defined by commutators of the basic tetrad vectors:
\begin{equation}
    \lambda_{abc}=(e_{a\mu,\nu}-e_{a\nu,\mu})e^\mu_{b}e^\nu_{c}.
\end{equation}
The only non-vanishing components of $\lambda_{abc}$ for the tetrad (\ref{tetrad}) read (up to antisymmetry):  
\begin{equation}\label{riccicoeffs}
    \begin{aligned}
    \lambda_{113}&=\frac{\alpha}{2}\bigg(\frac{A_2'}{A_2}-\frac{bA_{23}'}{\Sigma}\bigg),\;\;
    \lambda_{114}=-\frac{\beta}{2}\frac{aB_{23}'}{\Sigma},\;\;
    \lambda_{124}=\alpha\frac{bB_{23}'}{\Sigma},\;\;\\
    \lambda_{224}&=-\frac{\beta}{2}\bigg(\frac{B_2'}{B_2}+\frac{aB_{23}'}{\Sigma}\bigg),\;\;
    \lambda_{223}=-\frac{\alpha}{2}\frac{bA_{23}'}{\Sigma},\;\;
    \lambda_{213}=\beta\frac{aA_{23}'}{\Sigma},\\
    \lambda_{334}&=\frac{\beta}{2}\frac{aB_{23}'}{\Sigma},\;\;
    \lambda_{434}=\frac{\alpha}{2}\frac{bA_{23}'}{\Sigma}.\\
    \end{aligned}
\end{equation}

One can now find the following components of Ricci tensor using (\ref{RicciLL}): 
\begin{equation}\label{RicciTensor}
\begin{aligned}
R_{11}&=\frac{\left(\left(A_{23}'\right){}^2+\left(B_{23}'\right){}^2\right) \left(\alpha ^2 b^2-a^2 \beta ^2\right)}{2\Sigma ^2}-\frac{a \beta ^2 B_{23}''}{2 \Sigma }-\frac{a B_2' B_{23}'}{2 \Sigma ^2}-\frac{\alpha ^2 b
A_{23}''}{2 \Sigma }-\frac{b A_2' A_{23}'}{2 \Sigma ^2}+\frac{A_2''}{2 \Sigma },\\
R_{12}&=\frac{\sqrt{A_2} \sqrt{B_2} }{2 \Sigma ^2}\left(a A_{23}''+b B_{23}''\right),\\
R_{22}&=R_{11}-\frac{1}{2\Sigma}(A_2''+B_2''),\\
R_{33}&=\frac{a^2 \beta ^2 \left(\left(A_{23}'\right){}^2+\left(B_{23}'\right){}^2\right)}{2 \Sigma ^2}+\frac{a \beta ^2B_{23}''}{2 \Sigma }+\frac{a B_2' B_{23}'}{2 \Sigma ^2}-\frac{\alpha ^2 b A_{23}''}{2 \Sigma }+\frac{b A_2'
A_{23}'}{2 \Sigma ^2}-\frac{A_2''}{2 \Sigma },\\
R_{44}&=\frac{b^2 \alpha^2 \left(\left(A_{23}'\right){}^2+\left(B_{23}'\right){}^2\right)}{2 \Sigma ^2}+\frac{a \beta ^2B_{23}''}{2 \Sigma }-\frac{a B_2' B_{23}'}{2 \Sigma ^2}-\frac{\alpha ^2 b A_{23}''}{2 \Sigma }-\frac{b A_2'
A_{23}'}{2 \Sigma ^2}-\frac{B_2''}{2 \Sigma },\\
R_{13}&=R_{14}=R_{23}=R_{24}=R_{34}=0.
\end{aligned}
\end{equation}
\section{The action and equations of motion}
\subsection{Real form}
We consider the Einstein-Maxwell-dilaton-axion theory with a potential which is a consistent truncation of the four-dimensional ${\cal N}=4$ gauged supergravity with one Abelian vector field $A_{\mu}$. It is described by the action 
\begin{equation}\label{ActionMain}
    S=\frac{1}{16\pi}\int\bigg(-R+2\partial_\mu\phi \partial^\mu\phi +\frac{1}{2} e^{4\phi}\partial_\mu \kappa \partial^\mu \kappa +\frac{1}{l^2}V-e^{-2\phi}F_{\mu\nu}F^{\mu\nu} -\kappa F_{\mu\nu}\tilde{F}^{\mu\nu} \bigg)\sqrt{-g}d^4x,
\end{equation}
where $\tilde{F}^{\mu\nu}=\frac{1}{2}E^{\mu\nu\lambda\tau}F_{\lambda\tau}$, $E^{\mu\nu\lambda\tau}=\epsilon^{\mu\nu\lambda\tau}/\sqrt{-g}$, the quantity $l$ has dimension of length, and the dimensionless potential is 
\begin{equation} 
V=4+e^{-2\phi}+e^{2\phi}\left(\kappa^2+1\right).
\end{equation} 
Our convention for the Levi-Civita symbol is $\epsilon_{1234}=-\epsilon^{1234}=1,$ where 1234 stands for the sequence $t, \varphi, r,y$.

The corresponding equations of motion consist of the
modified Maxwell equations and Bianchi identities 
\begin{equation}\label{ModifiedMaxwell}
    \nabla_{\nu}(e^{-2\phi} F^{\mu\nu} + \kappa \tilde{F}^{\mu\nu})=0,
\end{equation}
\begin{equation}\label{ModifiedBianchi}
    \nabla_{\nu} \tilde{F}^{\mu\nu}=0,
\end{equation}
the dilaton equation with other fields acting as a source
\begin{equation}
\nabla_{\mu}\nabla^{\mu}\phi=\frac{1}{2}e^{-2\phi}F_{\mu\nu}F^{\mu\nu}+\frac{1}{2}(\nabla\kappa)^2-\frac{1}{2l^2}\left[e^{-2\phi}-e^{2\phi}(1+\kappa^2)\right],
\end{equation}
where $(\nabla\kappa)^2=g^{\mu\nu}(\nabla_\mu\kappa)(\nabla_\nu\kappa)$, and the axion   equation:
\begin{equation}
    \nabla^{\mu}(e^{4\phi}\nabla_{\mu}\kappa)=-F_{\mu\nu}\tilde{F}^{\mu\nu}+\frac{2}{l^2}\kappa e^{2\phi}.
\end{equation}
While the linear d'Alembert operator is separable
with our metric parametrization, its nonlinear versions are not. The same is true for Einstein's equations, which in our case are:
\begin{equation}\label{EinsteinEq}
R_{\mu\nu}=2\nabla_\mu\phi \nabla_\nu\phi+\frac{1}{2}e^{4\phi}\nabla_\mu\kappa \nabla_\nu\kappa+\frac{1}{2l^2}\big(4+e^{-2\phi}+e^{2\phi}(\kappa^2+1)\big)g_{\mu\nu}+e^{-2\phi}(2F_{\mu\lambda}F^{\lambda}_{\;\;\nu}+\frac{1}{2}g_{\mu\nu}F_{\lambda\tau}F^{\lambda\tau}).
\end{equation}

To obtain a solution to the entire nonlinear system of equations, we use the same strategy as Carter \cite{Carter:1968ks}: first, we extract from the system of equations some {\em linear} differential equations for the BF coefficients that reveal their polynomial structure. Then, the task of solving essentially nonlinear equations is reduced to finding relationships between the coefficients of all the polynomials involved, leaving some parameters free as physically significant.
\subsection{Complex form}
To implement this program, it is advantageous to rewrite the system in terms of a  complex 
axidilaton
field $z=\kappa+i e^{-2\phi}$. The action will take the form:
\begin{equation}\label{ActionAxidilaton}
    S=-\frac{1}{16\pi}\int\bigg(R+\frac{2\nabla z \nabla \Bar{z}}{(z-\Bar{z})^2}-\frac{1}{l^2}V-(iz\mathcal{F}_{\mu\nu}\mathcal{F}^{\mu\nu} + c.c.) \bigg)\sqrt{-g}d^4x,
\end{equation}
where $\mathcal{F}^{\mu\nu}=\frac{1}{2}\big(F^{\mu\nu}+i\tilde{F}^{\mu\nu}\big)$ and the potential will read
\begin{equation}
    V=4\bigg(1+\frac{i}{2}\frac{1+z\Bar{z}}{(z-\Bar{z})} \bigg).
\end{equation}
For $V=0$, the theory is   invariant under  $S$-duality transformations:  
\begin{equation}\label{S-duality}
    z\rightarrow\frac{a z + b}{c z + d},\;\;\;
    \mathcal{F}^{\mu\nu}\rightarrow (c \Bar{z} +d)\mathcal{F}^{\mu\nu},\;\;\;
    ad-bc=1.
\end{equation}
This leaves invariant the kinetic term of the axidilaton in (\ref{ActionAxidilaton}) and the full set of equations of motion, though not the action. The presence of the potential $V$ violates this invariance.

The tetrad components of the Einstein equations, following from the action \eqref{ActionAxidilaton} read:
\begin{equation}\label{EingaugedEMDA}
R_{ab}=T^{sc}_{ab}+\frac{(z-\bar{z})}{2i} T^{em}_{ab}+\frac{2}{l^2}\bigg(1+\frac{i}{2}\frac{1+z\bar{z}}{(z-\bar{z})} \bigg)\eta_{ab},
\end{equation}
where the reduced scalar term without trace part and potential is equal to
\begin{equation}\label{ScalarTem}
     T^{sc}_{ab}=-\frac{1}{(z-\Bar{z})^2}(z_{,a}\Bar{z}_{,b}+z_{,b}\Bar{z}_{a})
\end{equation}
(the derivatives with respect to tetrad vectors are  denoted by Latin indices after commas
$\partial_a=e_a^\mu\partial_\mu$), while the Maxwell term is standard:
\begin{equation}\label{Tem} T^{em}_{ab}=2\left(F_{ac}\tensor{F}{^{c}_b}+\frac{\eta_{ab}}4F_{cd}F^{cd}\right). 
\end{equation}
The reduced scalar stress tensor has nonzero components only in the $3,4$ sector (or $r,\,y$ in the coordinate basis). Note that on the left we use the Ricci tensor, not the Einstein tensor.

The first term on the right-hand side is invariant under $S$-duality transformations (\ref{S-duality}) by itself, which is easy to prove by a simple calculation. The invariance of the second term on the right-hand side is more difficult to verify. First, note that the axidilaton factor transforms as
\begin{equation}
    (z-\Bar{z})\rightarrow\frac{(z-\Bar{z})}{|cz+d|^2}.
\end{equation}
Then, for Einstein's equation to remain invariant, the Maxwell stress-energy tensor (\ref{Tem}) must transform as
\begin{equation} T^{em}_{ab}\rightarrow|cz+d|^2T^{em}_{ab},
\end{equation}
which is in fact the case and can be proved using the Schouten identity
\begin{equation}
    \tensor{\delta}{_\mu^{[\nu}}\tensor{\epsilon}{^{\rho \sigma \tau \lambda]}}=0.
\end{equation}
Thus, in absence of the potential $V$, Einstein equations are in fact invariant under the S-duality.

The second order equation  for the complex axidilaton field following from the action (\ref{ActionAxidilaton}) reads
\begin{equation}\label{AxidilatonEq0}
    \Box z-\frac{2\partial z \partial z}{(z-\Bar{z})}+\frac{i}{l^2}(z^2+1)-\frac{(z-\Bar{z})^2}{4}\big(iF_{\mu\nu}F^{\mu\nu}+F_{\mu\nu}\tilde{F}^{\mu\nu}\big)=0.
\end{equation}
\subsection{Maxwell field}
An ansatz for the vector one-form compatible with the metric parameterization (\ref{metricmain0}) was found by Carter \cite{Carter:1968ks} and is given by two functions $R(r)$ and $Y(y)$ as follows
\begin{equation}\label{Potential}
A=\frac{R}{\alpha\Sigma}e^1+\frac{Y}{\beta\Sigma}e^2.
\end{equation}
This one form has zero Lie derivatives with respect of two commuting Killing vectors.
It leads to separation of the Hamilton-Jacoby and Klein-Gordon equation for the charged particles.

Since the one-form depends only on two variables, the field strength two-form $F=dA$ has the components
\begin{equation}\label{fieldstrength}
    \begin{aligned}
        F_{13}=&\frac{1}{\Sigma}\bigg(-R'+R\bigg[\frac{A_2'}{2A_2} +\frac{bA_{23}'}{2\Sigma} - \frac{\lambda_{113}}{\alpha}\bigg] +\lambda_{213}\frac{Y}{\beta}\bigg),\\
        F_{24}=&\frac{1}{\Sigma}\bigg(-Y'+Y\bigg[\frac{B_2'}{2B_2} -\frac{aB_{23}'}{2\Sigma} + \frac{\lambda_{224}}{\beta}\bigg] -\lambda_{124}\frac{R}{\alpha}\bigg),\\
        F_{14}=&\frac{-R}{\alpha\Sigma}\bigg(\frac{a\beta B_{23}'}{2\Sigma}+\lambda_{114}\bigg),\\
        F_{23}=&\frac{Y}{\beta\Sigma}\bigg(\frac{\alpha b A_{23}'}{2\Sigma}+\lambda_{223}\bigg).\\
    \end{aligned}
\end{equation}
Using explicit expressions for the Ricci coefficients (\ref{riccicoeffs}) one finds that $F_{14}$ and $F_{23}$ vanish, so there are only two nonzero components of the field strength, are non-zero: $F_{13}$ and $F_{24}$, which also form the corresponding tetrad components of the Hodge-dual tensor: 
\begin{equation}\label{F}
    F_{13}=-\tilde{F}_{24}=
    \frac{A'_{23}(bR+aY)-\Sigma R'}{\Sigma^2},\quad
F_{24}=\tilde{F}_{13}=
    -\frac{B'_{23}(bR+aY)+\Sigma Y'} {\Sigma^2}
\end{equation}
Then  the Maxwell  energy-momentum tensor can be presented as follows:
\begin{equation}\label{Temcomponents}
    T^{em}_{11}= T^{em}_{22}=-T^{em}_{33}=T^{em}_{44}= \frac{\left(A_{23}' (a Y+b R)-\Sigma  R'\right){}^2+\left(B_{23}' (a Y+b R)+\Sigma  Y'\right){}^2}{\Sigma ^4}.
\end{equation}

Finally, the tetrad components of the modified Maxwell equations will take the form
\begin{equation}\label{MaxwellEq0}
    F^{ab}\partial_a(z-\Bar{z}) + (\nabla_\mu F^{\mu\nu})e^b_\nu (z-\Bar{z})+i\tilde{F}^{ab}\partial_a(z+\Bar{z})=0,
\end{equation}
where the divergence terms with account of antisymmetry of $F^{cb}$ and $\lambda_{abc}=-\lambda_{acb}$ read:
\begin{equation}
    (\nabla_\mu F^{\mu\nu})e^b_\nu=\tensor{F}{^{cb}_{,c}}+\frac{1}{2}F^{ca}(\tensor{\lambda}{_a^b_c}+\tensor{\lambda}{^b_{ca}}-\tensor{\lambda}{_{ca}^b})+
F^{cb}\tensor{\lambda}{^d_{dc}}.
\end{equation}

\section{Disentangling the equations}
\subsection{Extracting linear equations for BF coefficients}
 
Both the scalar and Maxwell energy-momentum tensors 
have zero mixed components $T_{12}$. So  the corresponding component of Einstein equations is vacuous, and using the  Ricci tensor component $R_{12}$  from the list 
(\ref{RicciTensor})
we obtain the following second order linear differential equation:
\begin{equation}\label{G12} aA_{23}''+bB_{23}''=0.
\end{equation}
Since the first term depends only on $r$, and the second only on $y$, it follows that the functions $A_{23},B_{23}$ must be polynomials of no greater than the second degree of the corresponding variables. Thus, the most general expression for these quantities is
\begin{equation}
\begin{aligned}
    A_{23}(r)&=\alpha_0+2 \alpha_1 r +  c r^2,\\
    B_{23}(y)&=- \beta_0 -2 \beta_1 y - \frac{a}{b}c y^2,
\end{aligned}
\end{equation}
where $\alpha_0,\beta_0,\alpha_1,\beta_1,c$ are  arbitrary constants.
The choice of gauge (\ref{ab}) still leaves room for constant shifts $r\to r+r_0,\,y\to y+y_0$, which can be used to set $\alpha_1=0,\;\beta_1 =0$ to simplify the calculations. Also, using the fact that the highest degrees of the polynomials $A_{23}$ and $B_{23}$ contain the same factor $c$, we can reset $c=1$ by rescaling $\varphi$. By also setting $b=1$ (by rescaling $t$), we get
\begin{equation}\label{A23B23}
\begin{aligned}
    A_{23}(r)&=\alpha_0+r^2,\\
    B_{23}(y)&=-\beta_0- ay^2,
\end{aligned}
\end{equation}
and therefore
\begin{equation}\label{Sigmaqu}
\Sigma=r^2+a^2y^2+\alpha_0+a\beta_0.
\end{equation}
 
The next quasilinear equation (becoming linear in absence of the scalar potential) can be obtained from the difference of $R_{11}-R_{22}$ components in  the Einstein's equations \eqref{EingaugedEMDA}. The reduced scalar stress tensor \eqref{ScalarTem} in this sector is zero, while contribution of the Maxwell stress-tensor in view of relations \eqref{Temcomponents} also vanishes, so we a left with potential terms only.  Then using the list \eqref{RicciTensor} one obtains
\begin{equation}\label{Ein11-Ein22}
A_2''+B_2''=\frac{8\Sigma}{l^2}\bigg(1+ \frac{i}{2} \frac{1+z \bar{z}}{z-\bar{z}} \bigg).
\end{equation}

The sum of the Einstein equations $R_{11}+R_{22}$ on the contrary does not contain the potential term and the reduced scalar stress tensor either, while $T^{em}_{11}+T^{em}_{22}\neq0$. In order to write the quantity $R_{11}+R_{22}$ in a convenient form let us introduce the function
\begin{equation}
    P=\frac{A_2-a^2B_2}{\Sigma},
\end{equation}
which will play the role of a "gravitational potential". Differentiating this definition twice on $r$ and $y$ one can express the second rerivatives of the Benenti coefficients
$A_2, B_2$ as follows
\begin{equation}
     \begin{aligned}
         A_2''&= \Sigma P_{,rr}  +\frac{2 a^2 B_2
\left(A_{23}'\right){}^2}{\Sigma ^2}-\frac{a^2 B_2A_{23}''}{\Sigma }-\frac{2 A_2\left(A_{23}'\right){}^2}{\Sigma ^2}+\frac{2 A_2'
A_{23}'}{\Sigma }+\frac{A_2 A_{23}''}{\Sigma
   },\\
   B_2''&=-\frac{1}{a^2}\Sigma P_{,yy} -\frac{2 a^2 B_2\left(B_{23}'\right){}^2}{\Sigma^2}+\frac{A_2
   B_{23}''}{a \Sigma }-\frac{2 a B_2' B_{23}'}{\Sigma}-\frac{a B_2 B_{23}''}{\Sigma }+\frac{2 A_2\left(B_{23}'\right){}^2}{\Sigma^2}.
     \end{aligned}
 \end{equation}
Substituting these relations back to the tetrad components $R_{11}$ and $R_{22}$ in    (\ref{RicciTensor}), one can see that all terms containing $A_i,B_i$ and their derivatives cancel out from $R_{11}+R_{22}$. As a result we obtain a quasilinear equation for $P$ with the source term
\begin{equation}\label{remainingEinstein}
    P_{,rr}+\frac{1}{a^2}P_{,yy}=-2i(z-\Bar{z})(F_{13}^2+F_{24}^2).
\end{equation}

\subsection{Ansatz for axidilaton}
For the following, we need a consistent ansatz for axidilaton, which has not been presented before. To extract a separate equation for $z$, we consider the  combinations of the Einstein equations components $R_{11}+R_{33}$ and
$R_{22}-R_{44}$.
In the first case contribution from the Maxwell tensor and the scalar potential in \eqref{EingaugedEMDA} vanish and  using the tetrad components $R_{11}$ from \eqref{RicciTensor} and the equation \eqref{G12} one arrives at
\begin{equation}\label{Ein11+Ein33}
    \frac{1}{2\Sigma^2}(4\Sigma-((A_{23}')^2+(B_{23}')^2))=\frac{2}{(z-\bar{z})^2}z_{,r}\bar{z}_{,r}.
\end{equation}

In the second case one obtains similarly
\begin{equation}\label{Ein22-Ein44}
    \frac{a^2}{2\Sigma^2}(4\Sigma-((A_{23}')^2+(B_{23}')^2))=\frac{2}{(z-\bar{z})^2}z_{,y}\bar{z}_{,y}.
\end{equation}
From these equations a separate axidilaton equation  follows:
\begin{equation}
    a^2 z_{,r}\Bar{z}_{,r}-z_{,y}\Bar{z}_{,y}=0.
\end{equation}
Another  equation for axidilaton follows from the Einstein equation  $R_{34}=0$ implying 
\begin{equation} \label{Ein34}
z_{,r}\bar{z}_{,y}+z_{,y}\bar{z}_{,r}=0.
\end{equation}
From these equations it follows that $z$ is a holomorphic or antiholomorphic function of $w=r+iay$, except at the poles.
Without loss of generality we choose the first option,
\begin{equation}
z=f(w),\quad w=r+iay.
\end{equation}
Then 
\begin{equation}\label{Zconditions1}       z_{,y}=iaz_{,r},
\end{equation}
and  the Laplace equation holds
\begin{equation}\label{Zconditions}
    z_{,yy}=-a^2z_{,rr}, 
\end{equation}
so we deal with a harmonic function.

A holomorphic function is needed, one that is single-valued in the entire complex plane and non-singular except at a simple pole. Such a function must be a fractional-linear transformation
\begin{equation}
    z=z_\infty \cdot\frac{r+iay+c_1}{r+iay+c_2},
\end{equation}
where $z_\infty,c_1,c_2$ are arbitrary complex constants.
It is clear that the constant $z_\infty$ fix the asymptotic values of the dilaton and the axion. For simplicity, we will assume here that $\phi_\infty=\kappa_\infty=0$,  that is,  $z_\infty=i$, and the axidilaton can ultimately be represented by the function:
\begin{equation}\label{axidilaton main function}
    z=i \frac{r+iay+c_1}{r+iay+c_2}.
\end{equation}
The remaining non-trivial axidilaton  equation \eqref{AxidilatonEq0} will read:
\begin{equation}\label{axidilatonEq1}
(A_2z_{,r})_{,r}+(B_2z _{,y})_{,y}-\frac{2}{(z-\Bar{z})}(A_2(z_{,r})^2+B_2(z_{,y})^2)-\frac{i}{l^2}(z^2+1)\Sigma=
    \frac{i(z-\Bar{z})^2\Sigma}{2}(F_{13}-iF_{24})^2.
\end{equation}

\subsection{Maxwell equations} 
In the system of modified Maxwell equations \eqref{MaxwellEq0}, by default only two equations are not satisfied, which have the form:
\begin{equation}\label{ModifiedMaxwell1}
    \alpha F_{13}(z-\Bar{z})_{,r}-(z-\Bar{z})\left[ F_{13} (\lambda_{223}-\lambda_{434})-F_{24} \lambda_{124}-\alpha F_{13,r}\right]+i\alpha F_{24}(z+\Bar{z})_{,r}=0,
\end{equation}
\begin{equation}\label{ModifiedMaxwell2}
    \beta F_{24}(z-\Bar{z})_{,y}+(z-\Bar{z})\left[F_{24} (\lambda_{114}-\lambda_{334})-F_{13} \lambda_{213}+\beta F_{24,y}\right]-i\beta F_{13}(z+\Bar{z})_{,y}=0.
\end{equation}
The components of the Maxwell tensor included here can be represented as
\begin{equation}\label{F1}
    F_{13}=
   -\frac{\partial}{\partial r}\bigg( \frac{R+aY}{\Sigma}\bigg),\quad
F_{24}=
    -\frac{1}{a}\frac{\partial}{\partial y}\bigg(\frac{R+aY}{\Sigma}\bigg),
\end{equation}
so the quantity in brackets plays the role of a "scalar potential".

It turns out that one can obtain simple linear equation on functions $R,\; Y$ from these modified Maxwell equations as follows. 
First, one has to multiply the equation \eqref{ModifiedMaxwell1} by $a/\alpha$ and than subtract it from the second equation \eqref{ModifiedMaxwell2} multiplied by $1/\beta$. Than taking into account the expressions for Ricci coefficients \eqref{riccicoeffs} (where we have already set $b=1$) one will obtain:
\begin{equation}
    (F_{24}-iF_{13})\big[(z+\bar{z})_{,y}-ia(z-\bar{z})_{,r}\big]-(z-\bar{z})\bigg[aF_{13,r}-F_{24,y}+\frac{2a}{\Sigma}(A'_{23}F_{13}+B'_{23}F_{24})\bigg]=0.
\end{equation}
The first term in this equation vanishes due to the condition \eqref{Zconditions1} so the equation reduces to 
\begin{equation}\label{preRY}
    aF_{13,r}-F_{24,y}+\frac{2a}{\Sigma}(A'_{23}F_{13}+B'_{23}F_{24})=0.
\end{equation}
Taking into account the expressions for tetrad components of the field strength \eqref{F}, one can obtain the following expression for the first two terms of this equation by a direct calculation
\begin{equation}\label{64}
    aF_{13,r}-F_{24,y}=\frac{1}{\Sigma^2}\Big[(aA''_{23}+B''_{23})(R+aY)-\Sigma(aR''-Y'')\Big]-\frac{2a}{\Sigma}(A'_{23}F_{13}+B'_{23}F_{24})
\end{equation}
Taking this relation and equation \eqref{G12} into account now, it becomes obvious that \eqref{preRY} and \eqref{64} give us a simple linear equation for the Maxwell functions of one variable:
\begin{equation}\label{RYconstraint}
    aR''-Y''=0.
\end{equation}
It follows from here that $R,\,Y$ are quadratic polynomials of  respective variables of which the highest coefficients are $a$- proportional. Then one can find a further simplification  from the fact that  
all our equations are invariant under the constant shift of the potential:
\begin{equation}\label{gauge}
    \frac{R+aY}{\Sigma}\rightarrow\frac{R+aY}{\Sigma}+C,
\end{equation}
where $C=$ const. Since $\Sigma$  given by the Eq. (\ref{Sigmaqu}) is also a quadratic polynomial with the ratio of leading coefficients equal to $a^2$, it is clear that by suitable choice of $C$ one can eliminate quadratic terms in $R$ and $Y$ simultaneously, so they can ultimately be chosen as the {\em linear} functions   of the respective variables:
\begin{equation} \label{MaxLin}
     R=R_0+q r,\qquad\qquad
         Y=Y_0-p y, 
\end{equation}
where $R_0,\, Y_0,\,q,\,p$ are real constants.
\section{Ungauged EMDA theory}
\subsection{Completing the integration}
In this section we find the general solution of the ungauged EMDA theory corresponding to the limit $l\rightarrow\infty$. The five-parameter family (plus two asymptotic values of the dilaton and axion, which we set to zero here) of rotating nutty dyons was obtained
in \cite{Galtsov:1994pd}
using Harrison transformations of the Kerr solution. Here, integrating the entire set of equations, we present a solution with two additional parameters.

In this limit the equations  \eqref{Ein11-Ein22} and  \eqref{axidilatonEq1},  simplify to   
\begin{equation}\label{EMDA Ein11-Ein22}
A''_2+B''_2=0,
\end{equation}
\begin{equation}\label{EMDA AxidilatonEq}
(A_2z_{,r})_{,r}+(B_2z _{,y})_{,y}-\frac{2}{(z-\Bar{z})}(A_2(z_{,r})^2+B_2(z_{,y})^2)=
    \frac{i(z-\Bar{z})^2\Sigma}{2}(F_{13}-iF_{24})^2.
\end{equation}
The first one implies that in addition to the previous polynomial parameterizations 
(\ref{A23B23},\ref{Sigmaqu},\ref{MaxLin}) one has
\begin{equation} \label{EMDA functions}
    A_{2}=a_0-2a_1 r + \lambda r^2,\qquad\qquad B_{2}=b_0+2b_1 y-\lambda y^2,    \end{equation}
where $a_i, b_i, \lambda $ are  real constants. 
It is seen that $g_{tt}= \lambda -2a/r$ as $r\to \infty$, so, assuming that spacetime is locally Minkowskian at spatial infinity, we must set $\lambda=1$, and $a_1=m$ - the Schwarzschild mass.

For the axidilaton, we take the fractional-linear function \eqref{axidilaton main function}. In total, at this stage, we have the fulfillment of part of the Einstein equations (\ref{G12}),(\ref{Ein34}),(\ref{EMDA Ein11-Ein22}) and one combination of two Maxwell equations (\ref{ModifiedMaxwell1})-(\ref{ModifiedMaxwell2}).
We will solve the remaining equations by fixing the relationships between the  unknown coefficients. From the equations (\ref{Ein11+Ein33}),(\ref{Ein22-Ein44}) we obtain the conditions 
\begin{equation}\label{GKconstraints1}
    \begin{aligned}
        c_1=&-c_2\equiv d,\\
        \alpha_0+a\beta_0=\frac{1}{2}&(c_1c_2^*+c_1^*c_2)=-dd^*\equiv\gamma_0,
    \end{aligned}
\end{equation}
so the axidilaton function simpifies to  
\begin{equation}\label{axidilatonGK2}
    z=i\frac{r+i a y + d}{r+i a y -d}.
\end{equation}

There are three more equations to solve, namely one of the modified Maxwell equations (\ref{ModifiedMaxwell1})-(\ref{ModifiedMaxwell2}), the Einstein equation (\ref{remainingEinstein}), and the axidilaton field equation (\ref{axidilatonEq1}). It is convenient to start with Maxwell's equations, since they do not depend on the functions $A_2, B_2$.
Taking into account (\ref{GKconstraints1}), we find that the only new constraint needed to satisfy the remaining Maxwell equation is
\begin{equation}\label{GKconstraints2}
d(q+ip)+d^*(q-ip)+2(R_0+aY_0)=0.
\end{equation}
Now using this condition together with (\ref{GKconstraints1}) and taking into account Einstein's equation (\ref{remainingEinstein}), we get two more constraints on our coefficients:
\begin{equation}\label{GKconstraints3}
    \begin{aligned}
        a_0-a^2& b_0 -(q^2+p^2)+ dd^*=0,\\
        d&=\frac{(q-ip)^2}{2(a_1 +i ab_1)}.
    \end{aligned}
\end{equation}
Finally, it can be verified that the remaining axidilaton equation \eqref{EMDA AxidilatonEq} is satisfied by substituting the \eqref{EMDA functions} together with the constraints (\ref{GKconstraints1}),(\ref{GKconstraints2}) and (\ref{GKconstraints3}). Thus, the entire system of equations of the ungauged EMDA theory is solved.
Defining now $b_0=\delta_1,\,\beta_0=\delta_0$, we finally obtain
\begin{equation}
    \begin{aligned}
        &A_{23}=r^2-|d|^2-a\delta_0,\\
        &B_{23}=-ay^2-\delta_0,
        \\
        &A_2= r^2-2a_1 r + (q^2+p^2)-|d|^2+a^2\delta_1,\\
        &B_2=- y^2+2b_1 y +\delta_1,\\ 
        &\Sigma=r^2+a^2 y^2 -|d|^2, \;\;\;\;  d=\frac{(q-ip)^2}{2(a_1 +i ab_1)},
    \end{aligned}
\end{equation}
where the domain of $y$ must be compact, so that $B_2$ is positive.
Note that the general solution was ``almost'' Minkowskian, up to fixing the constant
$\lambda=1$. We have thus proved that the general solution of EMDA theory admitting the Killing tensor is asymptotically flat (more precisely, locally flat, as we will see shortly). This correlates with recent uniqueness claims for static spacetimes admitting generalized photon surfaces \cite{Kobialko:2024rqr}, since
the existence of a slice-reducible Killing tensor (to which class our Killing tensor belongs) guarantees the existence of generalized photon surfaces \cite{Kobialko:2021aqg,Kobialko:2022ozq}.
So our current calculations hint that the above uniqueness theorem can be generalized to stationary solutions as well.
Previously, uniqueness statements for solutions of type D admitting the Killing-Yano tensor were presented in \cite{Houri:2007xz,Houri:2008ng,Krtous:2008tb,Frolov:2017kze}. It is worth noting again that our general solution is of Petrov type I and has no Killing-Yano structures.

\subsection{Physical nature of extra parameters}
Now that we have integrated the equations of motion, we need to establish the physical meaning of the constants of integration, but first we need to get rid of one unpleasant feature of the obtained solution. It turns out that if we consider, for example, the Kerr-Newman-NUT metric in coordinates where the linear terms in the functions $A_{23},\;B_{23}$ are absent, then it will have a singular static limit $a\rightarrow 0$, and this is what will happen to our solution too.
With these considerations in mind, we  make the coordinate shift  $y\rightarrow y+b_1$ and the following redefinitions of the parameters
$\delta_1\rightarrow\delta_1+b_1^2,\;\; \delta_0\rightarrow\delta_0+ab_1^2.$ Later on we use these new quantities.

To reveal physical meaning of the positive constant $\delta_1$, consider the equatorial plane 
$y=0$ at spatial infinity ($r\rightarrow \infty$). Two-dimensional line element than will take the form \begin{equation}\label{cosmic string}
dl^2_{(r,\varphi)}=\frac{\Sigma}{A_2}\bigg(dr^2+\frac{A_2}{\Sigma^2}\big[B_2 A_{23}^2-A_2 B_{23}^2\big]d\varphi^2\bigg)\rightarrow (dr^2+ \delta_1 r^2 d\varphi^2).
\end{equation}
Clearly the parameter $\delta_1$
introduces a conical singularity into the solution (the cosmic string)   \cite{Aryal:1986sz,Galtsov:1989ct,Hackmann:2010ir,Appels:2017xoe}. 
The cosmic string is removed by choosing $ \delta_1=1$.

To establish the physical meaning of the constants $ b_1,\delta_0$,
we consider the asymptotic behavior as $ r \to \infty$ of the rotation function $\omega$
\begin{equation}\label{omega}
\omega= -\frac{g_{t\varphi}}{g_{tt}}=\frac{A_2 B_{23}-a B_2 A_{23}}{A_2-a^2B_2}.
\end{equation}
We find that the NUT parameter $n$ corresponds to the setting $ b_1=n/a $:
\begin{equation}
\omega=B_{23}-aB_2=-2 n y-\delta_0-a\delta_1,
\end{equation}
and $ \delta_0+a\delta_1 $ is the parameter determining the position of the Misner strings. The north and south strings are symmetric when $\delta_0=-a\delta_1$. In this case, excluding also the cosmic string, we must set $\delta_1=1,\delta_0=-a$.

Now, if we define the complex dilaton charge using the asymptotics
\begin{equation}
z=i\bigg(1-\frac{2\mathcal{D}}{r} \bigg) + \mathcal{O}\bigg(\frac{1}{r^2}\bigg),
\end{equation}
and introduce the complex mass and electromagnetic charge as in \cite{Galtsov:1994pd}: $\mathcal{M}=m+i n,\;\mathcal{Q}=q+ip$, we obtain
\begin{equation}\label{ungauged Dilaton charge}
\mathcal{D}=-d=-\frac{(q-ip)^2}{2(m +i n)}\equiv -\frac{\mathcal{Q}^{*2}}{2\mathcal{M}}.
\end{equation}
Defining also the real charges  $e^2= |\mathcal{Q}|^2,\;D=|\mathcal{D}|=e^2/2\mu,\;\mu^2=m^2+n^2$, we can  rewrite  the solution  as
\begin{equation} \label{EMDA final solution}
    ds^2=\frac{A_2}{\Sigma}\big(dt + (a \cos^2\theta+2n\cos 
\theta-a)d\varphi\big)^2 - \frac{B_2}{\Sigma}\big(adt-(r^2-D^2+a^2+n^2)d\varphi\big)^2-\Sigma\bigg(\frac{dr^2}{A_2}+d\theta^2\bigg),
\end{equation}
where 
\begin{equation}\label{GK1}
    \begin{aligned}
        &A_2= r^2-2m r + e^2  -  D^2  +a^2-  n^2, \quad \\
        &B_2=\sin^2\theta,\\ 
        &\Sigma=r^2+(a\cos\theta +n)^2 -D^2,
    \end{aligned}
\end{equation}
where we set $y=\cos\theta$ to ensure $B_2>0$.
Now after transforming the coordinates $r\rightarrow r+r_{-}/2$ with $r_-=me^2/\mu^2$ in the solution of Ref. \cite{Galtsov:1994pd}, we find an exact match to our solution (\ref{EMDA final solution}-\ref{GK1}). 

Returning to our general solution, we conclude that two more parameters relative to the solution obtained using the Harrison transformations \cite{Galtsov:1994pd} are simply the possible conical defect and the Misner string asymmetry parameter.

For completeness, we also present the form of the vector potential and the axidilaton field in these coordinates. The potential one-form is given by the formula \eqref{Potential}, where the functions $R,\; Y$ read 
\begin{equation}\label{Ungauged RY}
\begin{aligned}
    R&=qr+\frac{1}{2}(\mathcal{D}(q+ip)+\mathcal{D}^*(q-ip))-np+a\delta_2,\\
    Y&=-py-\delta_2,
\end{aligned}
\end{equation}
where $\delta_2$ is insignificant real constant, dissapearing in the combination $R+aY$ defining  the Maxwell field tensor \eqref{F}. For axidilaton we have
\begin{equation}\label{Ungauged axidilaton}
    z=i\frac{r+iay+in-\mathcal{D}}{r+iay+in+\mathcal{D}},
\end{equation}
where the complex dilaton charge $\mathcal{D}$ is given by \eqref{ungauged Dilaton charge}.

\subsection{Ergosphere and horizon}
The Killing vector $\partial_t$ becomes null at the surfaces where $A_2-a^2 B_2=0$, which corresponds to    boundaries of the egoregion 
\begin{equation}\label{ergoregion}
    r_e^\pm= m\pm \sqrt{\mu^2- e^2+ D^2- a^2y^2}.
\end{equation}
The radii of the horizons $r_H^\pm$, satisfying the equation $A_2=0$,  are 
\begin{equation}\label{horizons}
    r_H^\pm= m \pm \sqrt{\mu^2- e^2+ D^2-a^2 }.
\end{equation}

Consider now the Killing vector $\xi=\partial_t+\Omega \partial_\varphi$ with some constant $\Omega$ that can still be timelike in the region $r^+_H<r<r^+_e$. The value of $\Omega$ on the surface $r=r_H^+$, where $\xi$ becomes null, will be equal to
\begin{equation}
    \Omega_H=\frac{a}{2\mu^2-e^2+2m\sqrt{\mu^2-e^2+D^2-  a^2}}.
\end{equation}
\subsection{Static limit}
Consider the case of a vanishing rotation parameter. The positions of the horizons are then simply given by
\begin{equation}  r_H^{\pm}=m\pm\frac{|2\mu^2-e^2|}{2\mu},
\end{equation}
and the position of a singularity is given by 
\begin{equation}
    r_s^2=D^2-n^2.
\end{equation}
To establish the relative position of these surfaces, we consider the value $(r_H^\pm)^2-r_s^2$, which after some algebraic transformations takes the form:
\begin{equation}\label{static limit rh-rs}
      (r_H^{\pm})^2-r_s^2=\frac{1}{\mu}\Big(\mu(2\mu^2-e^2)\pm m|2\mu^2-e^2|\Big).
 \end{equation}
First, we note that in the case where $2\mu^2=e^2$, which actually corresponds to the extremal limit, we obtain that the position of the singularity coincides with the position of the  horizon. Next, we should distinguish the following two cases:
\begin{itemize}
    \item $2\mu^2<e^2$.
Then we find that the ratio \eqref{static limit rh-rs} reduces to  
\begin{equation}
     (r_H^{\pm})^2-r_s^2=-\frac{1}{\mu}(\mu\mp m)(e^2-2\mu^2),
\end{equation}
so  the solution is a naked singularity, since $r_s$ is always larger than $r_H^+$.
\item $2\mu^2>e^2$. Then from \eqref{static limit rh-rs} we have the relation
\begin{equation}
(r_H^{\pm})^2-r_s^2=\frac{1}{\mu}(\mu\pm m)(2\mu^2-e^2),
\end{equation}
which tells us that for $n\neq 0$ the singularity is located under both horizons and coincides with the position of the inner horizon in the case of the vanishing NUT parameter  $n=0$.
\end{itemize}
\subsection{Absence of a wormhole}
The nutty solutions of Einstein-Maxwell theory
have a wormhole branch
\cite{Clement:2015aka,Clement:2022pjr} corresponding to everywhere regular metrics without horizons, so it is natural to look for a similar branch in EMDA theory.
From (\ref{horizons}) we see that if we want our solution to be horizonless, we must impose the following constraint on the parameters:
\begin{equation}
\mu^2-e^2+D^2-a^2<0,
\end{equation}
or, substituting 
$D=e^2/2\mu$,  we can rewrite this  as
\begin{equation}\label{horizon absence condition}
   2\mu(\mu-a)<e^2<2\mu(\mu +a).
\end{equation}
It is obvious that, unlike the Kerr-Newman-NUT solution, we cannot satisfy the no-horizon condition in EMDA theory in the static limit.
For $a\neq 0$
the charge of the horizonless solution can be presented as  
\begin{equation}\label{Q no horizon}
e^2=2\mu(\mu+\gamma a),\qquad \gamma\in(-1,1).
\end{equation}

For a solution to be a wormhole, we must also impose the condition that
$\Sigma$
is nonzero and positive definite everywhere. It is seen that in the case where $|n|<a$, there is always a ring singularity $r=D$ and $y=-n/a$, so we must consider $|n|>a$. In this case, a sufficient condition for $\Sigma$ to remain nonzero is
\begin{equation}\label{wormhole sing absence}
    |a\cos\theta-n|>D,
\end{equation}
for all values of $\theta\in[0,\pi]$.
Let us consider separately the following two   cases:
\begin{itemize}
    \item $n>0$. 
Then, given the connection between $D$ and $e^2$,   \eqref{wormhole sing absence} will be rewritten as 
\begin{equation}
    2\mu(n-a)>e^2.
\end{equation}
Using the no-horizon condition \eqref{Q no horizon}, we obtain
\begin{equation}
(n-a)>(\mu+\gamma a),
\end{equation}
which cannot be satisfied for any $\gamma$ in the range $(-1,1)$.
\item $n<0$.
In this case, the \eqref{wormhole sing absence} condition yields
\begin{equation}
2\mu(n+a)<-e^2,
\end{equation}
while
the no horizon condition now reduces to
\begin{equation}
(|n|-a)>(\mu+\gamma a),
\end{equation}
which also cannot be satisfied.
\end{itemize}
So we cannot satisfy the no-singularity condition together with the no-horizon condition. Thus, there is no parameter region where the solution can be a wormhole.
\subsection{Singularity}
As we have already mentioned, the position of the singularity is determined by the relation
\begin{equation}\label{sing circle}
    r^2+(a y+n)^2=D^2,
\end{equation}
which is actually an equation defining a circle of radius $D$ in the $r$-$y$ plane. The interior of this circle corresponds to $\Sigma<0$, so it is of no interest to us, while the exterior corresponds to the points where $\Sigma>0$ and the solution is regular. For convenience, we define a new coordinate $x=ay+n$, whose domain is $[x_{\rm min},x_{\rm max}]=[n-a,\;n+a]$. 
\begin{figure}
    \centering
\includegraphics[width=0.55\linewidth]{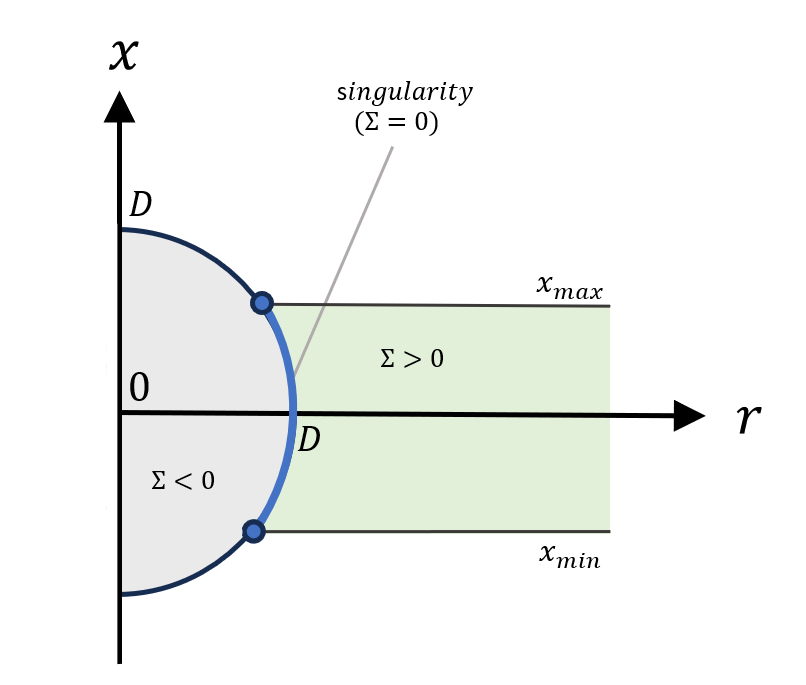}
 \caption{Position of a singularity.}
 \label{fig:enter-label}
\end{figure}

First, note that in the case $|n|>a$, one can obtain a condition on the electromagnetic charges $e$ that will guarantee the absence of a singularity at all, requiring that the interval $[x_{\rm min},x_{\rm max}]$ not intersect $[-D,D]$, which is impossible in the case $|n|<a$. One can obtain that, regardless of the sign of $n$, this condition is written as
\begin{equation}
    e^2<2\mu(|n|-a).
\end{equation}

Let us now consider the second case $|n|<a$. For this case, the singularity is represented by a part of the circle \eqref{sing circle} such that the point with $r_s=D$ must be included, since the interval $[x_{\rm min},x_{\rm max}]$ will contain the point $x=0$ (Fig. \ref{fig:enter-label}). Thus, the largest value of the radial coordinate corresponding to the singularity is $r_s=D$, and in order for the solution not to be a naked singularity, we must require the condition $r^+_H>r_s$, which can be written using the relation between the dilaton and the electromagnetic charge as
\begin{equation}\label{sing rhrs}
    \sqrt{(e^2-2\mu^2)^2-4a^2\mu^2}>e^2-2m\mu.
\end{equation}
If we consider the case $e^2>2m\mu$ than the square of this inequality leads to 
\begin{equation}
    (\mu-m)e^2+(a^2-n^2)\mu<0,
\end{equation}
which cannot be satisfied due to $|n|<a$, so we have to consider $e^2<2m\mu$, which satisfies the inequality \eqref{sing rhrs} by default. One more constraint following from the existence of the horizon is
\begin{equation}\label{sing rhrs1}
    e^2<2\mu(\mu-a).
\end{equation} 
One obtains, that in the case $|n|<a$ the relation $\mu-a<m$ must be fulfilled, so the condition for the existence of the horizon is enough for \eqref{sing rhrs} to be satisfied. Than in order for electromagnetic charge $e$ to remain real one also has to impose the condition $\mu-a>0$, which leads to 
\begin{equation}\label{sing rhrs2}
    m>\sqrt{a^2-n^2}.
\end{equation}
So in the case $|n|<a$ the conditions \eqref{sing rhrs1} and \eqref{sing rhrs2} guarantee the absence of a naked singularity.

\section{Gauged EMDA}
Now we come back to the full theory with a potential as described in Section 3.
\subsection{Forth order polynomials for $A_2,\; B_2$}
In the gauged case, the functions $A_{23},\;B_{23},\;R,\;Y$ can be considered exactly the same as in the previous section \eqref{EMDA functions}, and axidilaton will also take the general form \eqref{axidilaton main function}. But we need to clarify the nature of the BF functions $A_2,\;B_2$, since the right-hand side of Eq. \eqref{Ein11-Ein22} is now a nonlinear function of the variables $r$ and $y$. Therefore, we will leave these functions arbitrary for now and consider equations that are not affected by these functions.

From equations (\ref{Ein11+Ein33}) and (\ref{Ein22-Ein44}) we again obtain the constraints
\begin{equation}
\begin{aligned}
c_1&=-c_2\equiv d,\\
\alpha_0&+a\beta_0=-dd^*.
\end{aligned}
\end{equation}
We can rewrite this in another form by introducing a new real
constant $\delta_0$:
\begin{equation}
\begin{aligned}
\alpha_0&=-dd^*-\delta_0,\\
\beta_0&=\delta_0,
\end{aligned}
\end{equation}
Also, Maxwell's equations will again give us the same constraint as before, namely
\begin{equation}\label{Gaugeconstraints1}
d(q+ip)+d^*(q-ip)+2(R_0+aY_0)=0.
\end{equation}

Now, using the above constraints, we can see that the right-hand side of \eqref{Ein11-Ein22} is a second-order polynomial in $r$ and $y$ with separated variables, so $A_2$ and $B_2$ must now be fourth-order polynomials:
\begin{equation}
\begin{aligned}
&A_{2}=a_0 -2 a_1 r + a_2 r^2 +a_3 r^3 +a_4 r^4,\\
&B_{2}=b_0+ 2 b_1 y+b_2 y^2 +b_3 y^3 + b_4 y^4.\\
\end{aligned}
\end{equation}
Then, using these relations, from equation (\ref{Ein11-Ein22}) we get the following constraints
\begin{equation}\label{Gaugeconstraints2}
\begin{aligned}
a_4&=\frac{1}{l^2},\;\; b_4=\frac{a^2}{l^2},\\
a_3&=0, \;\;b_3=0,\\
a_2&+b_2= -\frac{2}{l^2} d d^*,
\end{aligned}
\end{equation}
therefore, we can set
\begin{equation}
\begin{aligned}
a_2&=-\frac{2}{l^2} |d|^2+\lambda,\\
b_2&=-\lambda,
\end{aligned}
\end{equation}
where $\lambda$ is an arbitrary real constant.

The remaining Einstein equation gives us
\begin{equation}\label{Gaugeconstraints3}
    \begin{aligned}
        a_0-a^2& b_0 -(q^2+p^2)-\frac{|d|^4}{l^2}+\lambda |d|^2=0,\\
        d&=\frac{(q-ip)^2}{2(a_1 +i a b_1)}.
    \end{aligned}
\end{equation}
So we have
\begin{equation}
    \begin{aligned}
        a_0&=q^2+p^2-\lambda |d|^2+\frac{|d|^4}{l^2}+a^2 \delta_1,\\
        b_0&=\delta_1.
    \end{aligned}
\end{equation}
The only equation which remains to be satisfied is the modified axidilaton equation (\ref{axidilatonEq1}), but it turns out that imposition of the constraints (\ref{Gaugeconstraints1}), (\ref{Gaugeconstraints2}) and (\ref{Gaugeconstraints3}) solves it, so the whole system of equations is satisfied.

Finally, we get the solution in the form
\begin{equation}\label{Gauged solution 1}
    \begin{aligned}
        &A_{23}=r^2-|d|^2-a\delta_0,\\
        &B_{23}=-a y^2 -\delta_0,\\
        &A_2=\frac{1}{l^2}r^4+\Big(\lambda-\frac{2}{l^2}|d|^2\Big)r^2-2a_1 r + (q^2+p^2)-\lambda |d|^2+a^2 \delta_1+\frac{|d|^4}{l^2},\\
        &B_2=\frac{a^2}{l^2}y^4-\lambda y^2+2 b_1 y + \delta_1,\\ 
        &\Sigma=r^2+a^2 y^2 -|d|^2, \;\;\;\;  d=\frac{(q-ip)^2}{2(a_1 +i a b_1)}.
    \end{aligned}
\end{equation}
\subsection{Physical nature of extra parameters}
In order to clarify the physical meaning of the free parameters, and  to establish that previously conjectured rotating solutions of the gauged EMDA theory are indeed special cases of our general solution \eqref{Gauged solution 1}, it is necessary to perform a coordinate transformation and a redefinition of the free parameters, as in the ungauged case. First, make the shift  $y \rightarrow y+y_0$ with some yet unfixed constant $y_0$, then make successive redefinitions
\begin{equation}
\begin{aligned}
&b_1 \rightarrow b_1-\frac{2 a^2 y_0^3}{l^2}+\lambda y_0,\\
&\lambda\rightarrow \lambda +\frac{a^2}{l^2}+\frac{6a^2y_0^2}{l^2},
\end{aligned}
\end{equation}
and then choose   $y_0=n/a$, and  $b_1=-2an/l^2$, where $n$ is the NUT parameter. Let us again introduce the complex dilaton charge $\mathcal{D}=-d$. Clearly the constant $a_1$ must be identified again with the mass $m$. After all   the final  solution will take the form:
\begin{equation}\label{Gauged final solution}
    ds^2=\frac{A_2}{\Sigma}\big(dt + (a y^2+2ny +\delta_0)d\varphi\big)^2 - \frac{B_2}{\Sigma}\big(adt-(r^2-|\mathcal{D}|^2-a\delta_0+n^2)d\varphi\big)^2-\Sigma\bigg(\frac{dr^2}{A_2}+\frac{dy^2}{B_2}\bigg),
\end{equation}
where
\begin{equation}\label{Gauged functions}
    \begin{aligned}
         &A_2=(r^2-|\mathcal{D}|^2)\bigg(\lambda+\frac{r^2-|\mathcal{D}|^2+a^2+6n^2}{l^2}\bigg)-2mr+(q^2+p^2)+a^2\delta_1-\lambda n^2+\frac{3n^2}{l^2}(a^2-n^2),\\
        &B_2=(1-y^2)\Big(\lambda-\frac{4an}{l^2}y-\frac{a^2}{l^2} y^2\Big)+\delta_1-\lambda,\\ 
        &\Sigma=r^2+(ay+n)^2 -|\mathcal{D}|^2, \;\;\;\;  \mathcal{D}=-\frac{(q-ip)^2}{2\big(m +i n\big[\lambda-\frac{a^2}{l^2}+\frac{4 n^2}{l^2}\big]\big)}.
    \end{aligned}
\end{equation}
In the limit $l\rightarrow\pm\infty$,  the solution (\ref{Gauged solution 1}) reduces to the solution of the ungauged theory (\ref{EMDA final solution}). Therefore the parameters $\delta_0,\; \delta_1$ will play the same role as in the ungauged EMDA.

Although the axidilaton charge $\mathcal{D}$ is related to the electromagnetic charges, we can see that formally considering the limit $\mathcal{D}\rightarrow 0$ with $q$ and $p$  unchanged and chosing $\lambda=1,\; \delta_1=1,\; \delta_0=-a$ in the solution \eqref{Gauged final solution} we arrive at Kerr-Newman-NUT-AdS solution \cite{Corral:2024lva}.
This could be considered more of a coincidence, since the theories are essentially different.

One important subcase is the Kerr-Sen-AdS solution, which appeared in some papers \cite{Wu:2020mby,Sakti:2022izj,Sakti:2022txd,Ali:2023ppg} without derivation   inspired by analogy the Kerr-Newman-AdS solution. Our rigorous derivation confirms the result in the special case of parameters $n=0,\;p=0,\; \mathcal{D}=-b=-q^2/2m,\; \lambda=1, \; \delta_1=1,\;\delta_0=-a$ and after coordinate transformation $r\rightarrow r+b$.

The axidilaton function and the functions $R,\;Y$, describing a potential one-form \eqref{Potential} are again given by Eqs. \eqref{Ungauged RY}-\eqref{Ungauged axidilaton}, where now the complex axidilaton charge is given by \eqref{Gauged functions}.

\subsection{Topological solutions}
For asymptotically AdS solutions one can expect occurence of different topologies  in the spirit of \cite{Lemos:1994xp,Cai:1996eg,Birmingham:1998nr,Vanzo:1997gw,Brill:1997mf}.
These can be identified as follows. First, note that $\lambda$ is no longer positive definite: one can safely consider $\lambda=0$ and $\lambda<0$. Correspondingly, the domain of $y$ (following from positivity of $B_2$)  is not necessarily limited to a finite region.
The last two cases just correspond to solutions with flat and hyperbolic topologies.
To see this, we rewrite the solution \eqref{Gauged final solution} as
\begin{equation}\label{Topological metric}
    ds^2=\frac{A_2-a^2B_2}{\Sigma}(dt-\omega d\varphi)^2-\frac{\Sigma}{A_2}dr^2-\Sigma d\sigma^2,
\end{equation}
where $\omega$ is defined  as in \eqref{omega}, and where we have  introduced a linear element $d\sigma^2$ of the 2-surface, parameterized by the coordinates $y, \; \varphi$ as follows
\begin{equation}\label{dsigma 1}
d\sigma^2=\frac{dy^2}{B_2}+ \frac{A_2B_2}{A_2-a^2 B_2}d\varphi^2.
 \end{equation}
Let us now consider this expression at spatial infinity $r\rightarrow\infty$ and in the case of the static limit $a\rightarrow0$. We obtain\begin{equation}\label{dsigma  2}    d\sigma^2=\frac{dy^2}{B_2}+ B_2d\varphi^2,
 \end{equation}
with $B_2=-\lambda y^2+\delta_1$. Having calculated the Gaussian curvature of this surface, we obtain
\begin{equation}
    R_\sigma=2\lambda,
\end{equation}
therefore we will be interested in three special cases of the constant $\lambda$, namely $\lambda=1,\;0,\;-1$ and the constant $\delta_1 =1$. Then we will get the following three cases

\begin{equation}
d\sigma^2 = 
 \begin{cases}
d\theta^2+\sin^2\theta d\varphi^2,  & \;\lambda=1,\;\;\;\; y=\cos\theta,\\
d\theta^2+d\varphi^2,  & \;\lambda=0, \;\;\;\;y=\theta,\\
 d\theta^2+\sinh^2\theta d\varphi^2,  & \;\lambda=-1,\; y=\cosh\theta.\\
 \end{cases}
\end{equation}
In presence of rotation $a\neq 0$, the Gaussian curvature $R_\sigma$ is no longer constant, as in the case of the static limit, but it turns out that these topological solutions can still be described in analogy with \cite{Klemm:1997ea}.
\subsection{Horizons}

Next we consider the \eqref{Gauged final solution} metric with spherical topology $\lambda=1$, remove the conical singularity $\delta_1=1$, establish symmetric Misner strings $\delta_0=-a$ and make a coordinate transformation $\varphi\rightarrow \varphi/\Xi\;$, obtaining
\begin{equation}\label{Gauged metric horizons}
    \begin{aligned}
         &A_2=(r^2-|\mathcal{D}|^2)\bigg(1+\frac{r^2-|\mathcal{D}|^2+a^2+6n^2}{l^2}\bigg)-2mr+(q^2+p^2)+\bigg(1+\frac{3n^2}{l^2}\bigg)(a^2-n^2),\\
        &B_2=\sin^2\theta\bigg(1-\frac{4an}{l^2}\cos\theta-\frac{a^2}{l^2} \cos^2\theta\bigg),\\ 
        &A_{23}=\frac{r^2-|\mathcal{D}|^2+a^2+n^2}{\Xi},\\
        &B_{23}=\frac{a-2n\cos\theta-a \cos^2\theta }{\Xi},\\
        &\Sigma=r^2+(a\cos \theta+n)^2 -|\mathcal{D}|^2, \;\;\;\;  \mathcal{D}=-\frac{(q-ip)^2}{2\big(m +i n\big[1-\frac{a^2}{l^2}+\frac{4 n^2}{l^2}\big]\big)},\;\;\;\Xi=1-\frac{a^2}{l^2}.
    \end{aligned}
\end{equation}
The angular velocity of locally non-rotating observers 
\begin{equation}
    \Omega=-\frac{g_{t\varphi}}{g_{\varphi \varphi}}=\frac{A_2B_{23}-aB_2 A_{23}}{A_2 (B_{23})^2-a\sin^2\theta (A_{23})^2}
\end{equation}
at spatial infinity $r\rightarrow \infty$  reduces to
\begin{equation}
    \Omega_{\infty}=-\frac{a}{l^2},
\end{equation}
which means that the asymotitic is rotating. The corresponding value at the horizon $r=r_+$, $A_2(r_+)=0$, equal to
\begin{equation}
\Omega_{H}=\frac{a\Xi}{r_+^2+a^2+n^2-|\mathcal{D}|^2},
\end{equation}
can be interpreted as the angular velocity of the horizon relative to a rotating frame. The angular velocity of the horizon relative to a frame  reference static at infinity can be defined as
\begin{equation}
    \Omega_H-\Omega_\infty=\frac{a}{r_+^2+a^2+n^2-|\mathcal{D}|^2}\bigg( 1+\frac{r_+^2+n^2-|\mathcal{D}|^2}{l^2}\bigg).
\end{equation}

Let us now consider the fourth-degree equation $A_2=0$, the roots of which determine the position of the horizons. This equation can be written as
\begin{equation}\label{quartic}
r^4+a_2r^2+a_1r+a_0=0,
\end{equation}
where we have defined
\begin{equation}
    \begin{aligned}
        a_2&= l^2 +a^2 -2|\mathcal{D}|^2+6n^2,\\
        a_1&=-2ml^2,\\
a_0&=l^2|\mathcal{Q}|^2+|\mathcal{D}|^4+(3n^2+l^2)(a^2- n^2)-(l^2 +a^2 +6n^2)|\mathcal{D}|^2.
    \end{aligned}
\end{equation}

It turns out that we can find the extremal value of the mass parameter $m_{\rm{ext}}$ and the location of the extremal event horizon $r^{\rm{ext}}_H$ without solving the quartic equation \eqref{quartic}. For $m=m_{\rm{ext}}$, the outer event horizon $r_+$ and the inner Cauchy horizon $r_-$ coincide, so the solution is an extremal AdS black hole. According to the Ferrari formula, the positive roots of our quartic equation can be written as
\begin{equation}\label{Ferrari}
    r_\pm=\frac{1}{2}\bigg(\sqrt{y_1-a_2}\pm\sqrt{2\sqrt{y_1^2-4a_0}-(y_1+a_2)}\bigg),
\end{equation}
 where $y_1$ is a real root of the resolvent cubic equation
\begin{equation}\label{Resolvent}
    y^3-a_2y^2-4a_0y-(a_1^2-4a_0a_2)=0.
\end{equation}
From \eqref{Ferrari} we can obtain the extremality condition on $y_1$, namely
\begin{equation}
y_1=\frac{1}{3}(a_2+2\sqrt{(a_2)^2+12a_0}),
\end{equation}
now substituting this relation into the equation of the resolvent \eqref{Resolvent} and considering it as an equation for the mass, we  obtain
\begin{equation}\label{AdS mass extremal}
    m_{\rm{ext}}=\frac{l}{3\sqrt{6}}\bigg(\eta+2\Big(1+\frac{a^2}{l^2}\Big)-\frac{4|\mathcal{D}|^2}{l^2}+\frac{12n^2}{l^2}\bigg)\cdot\bigg(\eta-\Big(1+\frac{a^2}{l^2}\Big)+\frac{2|\mathcal{D}|^2}{l^2}-\frac{6n^2}{l^2}\bigg)^{1/2},
\end{equation}
where
\begin{equation}
    \eta=\sqrt{\bigg(1+\frac{a^2}{l^2}\bigg)^2+\frac{12}{l^2}\bigg(a^2+|\mathcal{Q}|^2-\frac{4}{3}|\mathcal{D}|^2\bigg)-\frac{16}{l^4}\bigg[a^2\big(|\mathcal{D}|^2-3n^2\big)-|\mathcal{D}|^2\big(|\mathcal{D}|^2-6n^2\big)\bigg]}.
\end{equation}
And the extremal value of the horizon radius than reads 
\begin{equation}
    r^{\rm{ext}}_H=\frac{l}{\sqrt{6}}\bigg(\eta-1-\frac{a^2}{l^2}+2|\mathcal{D}|^2-6n^2\bigg)^{1/2}.
\end{equation}

We can now again formally consider the limit $|\mathcal{D}|\rightarrow0$ with nonvanishing electromagnetic charges. One finds, that in this case also putting $n=0$ the quantities $m_{\rm{ext}}$ and $r^{\rm{ext}}_H$ reduce to those of the Kerr-Newman-AdS solution \cite{Caldarelli:1999xj,Aliev:2007qi}.

 \section {Conclusions} 
We performed a full integration of the equations of motion of the ungauged and gauged EMDA theory, which is a truncated version of ${\cal N}=4$ supergravity with one vector field on the class of stationary axisymmetric spacetimes admitting a second-rank Killing tensor. Previously, similar calculations were restricted to the Einstein-Maxwell theory and type D metrics. We used Carter's approach, recently extended to algebraically non-special spacetimes and the presence of scalar fields. Special attention was paid to the dilaton and axion fields, which were not previously considered in this approach. We were able to extract from the full coupled nonlinear system of equations a pair of sepaate equations for the complex axidilaton, whose analytical structure implies the choice of the corresponding ansatz in the form of a fractional linear function. Then, separate equations were derived for the four most important BF coefficient functions, implying their polynomial structure. This was possible both in the ungauged EMDA theory and in the gauged theory with potential. Then the problem of integrating   nonlinear systems of equations was reduced to establishing relationships between the coefficients of the polynomials.

In the ungauged case, it was shown that the general solution is necessarily asymptotically flat or locally flat. It contains two more parameters than the nutty EMDA dyon found earlier by the Harrison transformations: one is a conical parameter, and the other determines the Misner string configuration. For the general set of parameters, the solution is either a black hole or a naked singularity, and does not contain a wormhole branch like the Kerr-Newman-NUT solution of Einstein-Maxwell theory.

In the gauged case, the general solution has AdS asymptotics and is endowed with a mass, a NUT parameter, electric and magnetic charges, a cone parameter, a Misner string parameter, and a discrete topological parameter defining the spherical, flat and hyperbolic topologies of the two-dimensional section. Such a solution has never been derived analytically before, although a special case without derivation, known as the Kerr-Sen-AdS metric, has been proposed. This metric has been rigorously verified and extended with new parameters.
 
This approach can be applied to other four-dimensional ungauged and gauged supergravities.
\begin{acknowledgments}
The authors thank G\'erard Cl\'ement, Kirill Kobialko, Alexander Kulitskii and Alexander Alexeev for valuable suggestions
and discussions.  The work of D.G. was supported by the Foundation for the Advancement of Theoretical
Physics and Mathematics 'BASIS'. \end{acknowledgments}


\begin{thebibliography}{20}
%\cite{Ernst:1967wx}
\bibitem{Ernst:1967wx}
F.~J.~Ernst,
``New formulation of the axially symmetric gravitational field problem,''
Phys. Rev. \textbf{167}, 1175-1179 (1968).
%doi:10.1103/PhysRev.167.1175
%525 citations counted in INSPIRE as of 26 Feb 2025, 

%\cite{Ernst:1967by}
\bibitem{Ernst:1967by}
F.~J.~Ernst,
``New Formulation of the Axially Symmetric Gravitational Field Problem. II,''
Phys. Rev. \textbf{168}, 1415-1417 (1968).
%doi:10.1103/PhysRev.168.1415
%370 citations counted in INSPIRE as of 26 Feb 2025

\bibitem{Neugebauer:1969wr}
G.~Neugebauer and D.~Kramer,
``A method for the construction of stationary einstein-maxwell fields. (in german),''
Annalen Phys. \textbf{24}, 62-71 (1969).
%78 citations counted in INSPIRE as of 26 Feb 2025G. 
\bibitem{Neugebauer}
G.~Neugebauer, "Untersuchungen zu
Einstein-Maxwell-Feldern mit eindimensionaler Bewegungsgruppe",
Habilitationsschrift, FSU Jena (1969).


%\cite{Maison:1979kx}
\bibitem{Maison:1979kx}
D.~Maison,
``Ehlers-Harrison type transformations for Jordan's extended theory of gravitation ''
Gen. Rel. Grav. \textbf{10}, 717-723 (1979).
%doi:10.1007/BF00756907
%122 citations counted in INSPIRE as of 26 Feb 2025

%\cite{Clement:1986kzn}
\bibitem{Clement:1986kzn}
G.~Clement,
``Stationary Solutions in Five-dimensional General Relativity,''
Gen. Rel. Grav. \textbf{18}, 137-160 (1986).
%IPUC-84-3-ADD.
%33 citations counted in INSPIRE as of 01 Mar 2025

%\cite{Breitenlohner:1987dg}
\bibitem{Breitenlohner:1987dg}
P.~Breitenlohner, D.~Maison and G.~W.~Gibbons,
``Four-Dimensional Black Holes from Kaluza-Klein Theories,''
Commun. Math. Phys. \textbf{120} (1988), 295.
%doi:10.1007/BF01217967
%407 citations counted in INSPIRE as of 20 Sep 2024

%\cite{Galtsov:1994pd}
\bibitem{Galtsov:1994pd}
D.~V.~Galtsov and O.~V.~Kechkin,
``Ehlers-Harrison type transformations in dilaton - axion gravity,''
Phys. Rev. D \textbf{50} (1994), 7394-7399
%doi:10.1103/PhysRevD.50.7394
[arXiv:hep-th/9407155 [hep-th]].
%108 citations counted in INSPIRE as of 20 Sep 2024

%\cite{Galtsov:1994sjr}
\bibitem{Galtsov:1994sjr}
D.~V.~Gal'tsov,
``Integrable systems in stringy gravity,''
Phys. Rev. Lett. \textbf{74}, 2863-2866 (1995)
%doi:10.1103/PhysRevLett.74.2863
[arXiv:hep-th/9410217 [hep-th]].
%72 citations counted in INSPIRE as of 26 Feb 2025

%\cite{Breitenlohner:1998cv}
\bibitem{Breitenlohner:1998cv}
P.~Breitenlohner and D.~Maison,
``On nonlinear sigma models arising in (super-)gravity,''
Commun. Math. Phys. \textbf{209} (2000), 785-810
%doi:10.1007/s002200050038
[arXiv:gr-qc/9806002 [gr-qc]].
%37 citations counted in INSPIRE as of 20 Sep 2024

%\cite{Youm:1997hw}
\bibitem{Youm:1997hw}
D.~Youm,
``Black holes and solitons in string theory,''
Phys. Rept. \textbf{316} (1999), 1-232
%doi:10.1016/S0370-1573(99)00037-X
[arXiv:hep-th/9710046 [hep-th]].
%219 citations counted in INSPIRE as of 20 Sep 2024

%\cite{Bouchareb:2007ax}
\bibitem{Bouchareb:2007ax}
A.~Bouchareb, G.~Clement, C.~M.~Chen, D.~V.~Gal'tsov, N.~G.~Scherbluk and T.~Wolf,
``G(2) generating technique for minimal D=5 supergravity and black rings,''
Phys. Rev. D \textbf{76} (2007), 104032
[erratum: Phys. Rev. D \textbf{78} (2008), 029901]
%doi:10.1103/PhysRevD.76.104032
[arXiv:0708.2361 [hep-th]].
%76 citations counted in INSPIRE as of 20 Sep 2024

%\cite{Chow:2013tia}\cite{Chow:2014cca}
\bibitem{Chow:2013tia}
D.~D.~K.~Chow and G.~Comp\`ere,
``Seed for general rotating non-extremal black holes of $\mathcal {N}= 8$ supergravity,''
Class. Quant. Grav. \textbf{31}, 022001 (2014)
%doi:10.1088/0264-9381/31/2/022001
[arXiv:1310.1925 [hep-th]].
%64 citations counted in INSPIRE as of 15 Jan 2025

%\cite{Chow:2014cca}
\bibitem{Chow:2014cca}
D.~D.~K.~Chow and G.~Comp\`ere,
``Black holes in N=8 supergravity from SO(4,4) hidden symmetries,''
Phys. Rev. D \textbf{90}, no.2, 025029 (2014)
%doi:10.1103/PhysRevD.90.025029
[arXiv:1404.2602 [hep-th]].
%75 citations counted in INSPIRE as of 15 Jan 2025

%\cite{Bogush:2020obx}
\bibitem{Bogush:2020obx}
I.~Bogush, G.~Cl\'ement, D.~Gal'tsov and D.~Torbunov,
``Nutty Kaluza-Klein dyons revisited,''
Phys. Rev. D \textbf{103} (2021) no.6, 064045
%doi:10.1103/PhysRevD.103.064045
[arXiv:2009.07922 [gr-qc]].
%4 citations counted in INSPIRE as of 20 Sep 2024

%\cite{Chong:2004na} 
\bibitem{Chong:2004na}
Z.~W.~Chong, M.~Cvetic, H.~Lu and C.~N.~Pope,
``Charged rotating black holes in four-dimensional gauged and ungauged supergravities,''
Nucl. Phys. B \textbf{717}, 246-271 (2005)
%doi:10.1016/j.nuclphysb.2005.03.034
[arXiv:hep-th/0411045 [hep-th]].
%146 citations counted in INSPIRE as of 18 Jan 2025

%\cite{Chow:2010sf}
\bibitem{Chow:2010sf}
D.~D.~K.~Chow,
``Single-charge rotating black holes in four-dimensional gauged supergravity,''
Class. Quant. Grav. \textbf{28}, 032001 (2011)
%doi:10.1088/0264-9381/28/3/032001
[arXiv:1011.2202 [hep-th]].

%\cite{Chow:2010fw}
\bibitem{Chow:2010fw}
D.~D.~K.~Chow,
``Two-charge rotating black holes in four-dimensional gauged supergravity,''
Class. Quant. Grav. \textbf{28}, 175004 (2011)
%doi:10.1088/0264-9381/28/17/175004
[arXiv:1012.1851 [hep-th]].
%\cite{Klemm:2012vm}

\bibitem{Klemm:2012vm}
D.~Klemm and O.~Vaughan,
``Nonextremal black holes in gauged supergravity and the real formulation of special geometry II,''
Class. Quant. Grav. \textbf{30}, 065003 (2013)
%doi:10.1088/0264-9381/30/6/065003
[arXiv:1211.1618 [hep-th]].
%44 citations counted in INSPIRE as of 18 Jan 2025

 

%\cite{Gnecchi:2013mja}
\bibitem{Gnecchi:2013mja}
A.~Gnecchi, K.~Hristov, D.~Klemm, C.~Toldo and O.~Vaughan,
``Rotating black holes in 4d gauged supergravity,''
JHEP \textbf{01}, 127 (2014)
%doi:10.1007/JHEP01(2014)127
[arXiv:1311.1795 [hep-th]].
%103 citations counted in INSPIRE as of 15 Jan 2025

\bibitem{Chow:2013gba}
D.~D.~K.~Chow and G.~Comp\`ere,
``Dyonic AdS black holes in maximal gauged supergravity,''
Phys. Rev. D \textbf{89}, no.6, 065003 (2014)
%doi:10.1103/PhysRevD.89.065003
[arXiv:1311.1204 [hep-th]].
%116 citations counted in INSPIRE as of 15 Jan 2025

%\cite{Zhu:2024jhw}
\bibitem{Zhu:2024jhw}
X.~D.~Zhu, D.~Wu and D.~Wen,
``Topological classes of thermodynamics of the rotating charged AdS black holes in gauged supergravities,''
Phys. Lett. B \textbf{856}, 138919 (2024)
%doi:10.1016/j.physletb.2024.138919
[arXiv:2402.15531 [hep-th]].
%24 citations counted in INSPIRE as of 18 Jan 2025

%\cite{Wu:2020cgf}\cite{Gallerati:2021cty}\cite{Anabalon:2024cnb}\cite{Zhu:2024jhw}
\bibitem{Anabalon:2024cnb}
A.~Anabal\'on, S.~Maurelli, M.~Oyarzo and M.~Trigiante,
``The Instability of Low-Temperature Black Holes in Gauged $\mathcal{N}=8$ Supergravity,''
[arXiv:2411.09454 [hep-th]].
%0 citations counted in INSPIRE as of 18 Jan 2025

%\cite{Wu:2020cgf}
\bibitem{Wu:2020cgf}
D.~Wu, P.~Wu, H.~Yu and S.~Q.~Wu,
``Are ultraspinning Kerr-Sen- AdS$_4$ black holes always superentropic?,''
Phys. Rev. D \textbf{102}, no.4, 044007 (2020)
%doi:10.1103/PhysRevD.102.044007
[arXiv:2007.02224 [gr-qc]].
%42 citations counted in INSPIRE as of 15 Jan 2025

%\cite{Gallerati:2021cty}
\bibitem{Gallerati:2021cty}
A.~Gallerati,
``New Black Hole Solutions in $N = 2$ and $N = 8$ Gauged Supergravity,''
Universe \textbf{7}, no.6, 187 (2021).
%doi:10.3390/universe7060187
%8 citations counted in INSPIRE as of 15 Jan 2025

%\cite{Carter:1968ks}
\bibitem{Carter:1968ks}
B.~Carter,
``Hamilton-Jacobi and Schrodinger separable solutions of Einstein's equations,''
Commun. Math. Phys. \textbf{10} (1968) no.4, 280-310.
%doi:10.1007/BF03399503
%827 citations counted in INSPIRE as of 20 Sep 2024

%\cite{Plebanski:1975xfb}
\bibitem{Plebanski:1975xfb}
J.~F.~Pleba\~nski,
``A class of solutions of Einstein-Maxwell equations,''
Annals Phys. \textbf{90}, no.1, 196-255 (1975).
%doi:10.1016/0003-4916(75)90145-1
%120 citations counted in INSPIRE as of 19 Jan 2025

%\cite{Frolov:2017kze}
\bibitem{Frolov:2017kze}
V.~P.~Frolov, P.~Krtous and D.~Kubiznak,
``Black holes, hidden symmetries, and complete integrability,''
Living Rev. Rel. \textbf{20}, no.1, 6 (2017)
%doi:10.1007/s41114-017-0009-9
[arXiv:1705.05482 [gr-qc]].
%197 citations counted in INSPIRE as of 24 Jan 2025

%\cite{Galtsov:2024vqo}
\bibitem{Galtsov:2024vqo}
D.~Gal'tsov and A.~Kulitskii,
``Petrov types, separability, and generalized photon surfaces of supergravity black holes,''
Phys. Rev. D \textbf{110} (2024) no.12, 124008
%doi:10.1103/PhysRevD.110.124008
[arXiv:2409.13324 [gr-qc]].
%1 citations counted in INSPIRE as of 25 Dec 2024

%\cite{Benenti:1979erw}
\bibitem{Benenti:1979erw}
S.~Benenti and M.~Francaviglia,
``Remarks on certain separability structures and their applications to general relativity,''
Gen. Rel. Grav. \textbf{10} (1979) no.1, 79-92.
%doi:10.1007/bf00757025
%82 citations counted in INSPIRE as of 20 Sep 2024

%\cite{Anabalon:2016hxg}
\bibitem{Anabalon:2016hxg}
A.~Anabalon and C.~Batista,
``A Class of Integrable Metrics,''
Phys. Rev. D \textbf{93}, no.6, 064079 (2016)
%doi:10.1103/PhysRevD.93.064079
[arXiv:1602.02037 [gr-qc]].
%6 citations counted in INSPIRE as of 18 Jan 202
%\cite{Wu:2020mby,Sakti:2022izj,Sakti:2022txd,Ali:2023ppg}
\bibitem{Wu:2020mby}
D.~Wu, S.~Q.~Wu, P.~Wu and H.~Yu,
``Aspects of the dyonic Kerr-Sen- AdS$_4$ black hole and its ultraspinning version,''
Phys. Rev. D \textbf{103}, no.4, 044014 (2021)
%doi:10.1103/PhysRevD.103.044014
[arXiv:2010.13518 [gr-qc]].
%31 citations counted in INSPIRE as of 15 Jan 2025

%\cite{Sakti:2022izj}
\bibitem{Sakti:2022izj}
M.~F.~A.~R.~Sakti and P.~Burikham,
``Dual CFT on a dyonic Kerr-Sen black hole and its gauged and ultraspinning counterparts,''
Phys. Rev. D \textbf{106}, no.10, 106006 (2022)
%doi:10.1103/PhysRevD.106.106006
[arXiv:2206.10868 [hep-th]].
%9 citations counted in INSPIRE as of 18 Jan 2025

%\cite{Sakti:2022txd}
\bibitem{Sakti:2022txd}
M.~F.~A.~R.~Sakti,
``Hidden conformal symmetry for dyonic Kerr-Sen black hole and its gauged family,''
Eur. Phys. J. C \textbf{83}, no.3, 255 (2023)
%doi:10.1140/epjc/s10052-023-11412-2
[arXiv:2208.02722 [hep-th]].
%6 citations counted in INSPIRE as of 01 Mar 2025
%\cite{Ali:2023ppg}
\bibitem{Ali:2023ppg}
M.~S.~Ali, S.~G.~Ghosh and A.~Wang,
``Thermodynamics of Kerr-Sen-AdS black holes in the restricted phase space,''
Phys. Rev. D \textbf{108}, no.4, 044045 (2023)
%doi:10.1103/PhysRevD.108.044045
[arXiv:2308.00489 [gr-qc]].
%14 citations counted in INSPIRE as of 26 Feb 2025

%\cite{Lemos:1994xp,Cai:1996eg,Birmingham:1998nr,Vanzo:1997gw,Brill:1997mf}\cite{Gnecchi:2013mja}\cite{Zhu:2024jhw}
\bibitem{Lemos:1994xp}
J.~P.~S.~Lemos,
``Cylindrical black hole in general relativity,''
Phys. Lett. B \textbf{353}, 46-51 (1995)
%doi:10.1016/0370-2693(95)00533-Q
[arXiv:gr-qc/9404041 [gr-qc]].
%420 citations counted in INSPIRE as of 19 Jan 2025

%\cite{Cai:1996eg}
\bibitem{Cai:1996eg}
R.~G.~Cai and Y.~Z.~Zhang,
``Black plane solutions in four-dimensional space-times,''
Phys. Rev. D \textbf{54}, 4891-4898 (1996)
%doi:10.1103/PhysRevD.54.4891
[arXiv:gr-qc/9609065 [gr-qc]].
%303 citations counted in INSPIRE as of 19 Jan 2025

%\cite{Birmingham:1998nr}
\bibitem{Birmingham:1998nr}
D.~Birmingham,
``Topological black holes in Anti-de Sitter space,''
Class. Quant. Grav. \textbf{16}, 1197-1205 (1999)
%doi:10.1088/0264-9381/16/4/009
[arXiv:hep-th/9808032 [hep-th]].
%479 citations counted in INSPIRE as of 19 Jan 2025

%\cite{Vanzo:1997gw}
\bibitem{Vanzo:1997gw}
L.~Vanzo,
``Black holes with unusual topology,''
Phys. Rev. D \textbf{56}, 6475-6483 (1997)
%doi:10.1103/PhysRevD.56.6475
[arXiv:gr-qc/9705004 [gr-qc]].
%366 citations counted in INSPIRE as of 19 Jan 2025

%\cite{Brill:1997mf}
\bibitem{Brill:1997mf}
D.~R.~Brill, J.~Louko and P.~Peldan,
``Thermodynamics of (3+1)-dimensional black holes with toroidal or higher genus horizons,''
Phys. Rev. D \textbf{56}, 3600-3610 (1997)
%doi:10.1103/PhysRevD.56.3600
[arXiv:gr-qc/9705012 [gr-qc]].
%322 citations counted in INSPIRE as of 19 Jan 2025

%\cite{Klemm:1997ea}
\bibitem{Klemm:1997ea}
D.~Klemm, V.~Moretti and L.~Vanzo,
``Rotating topological black holes,''
Phys. Rev. D \textbf{57}, 6127-6137 (1998)
[erratum: Phys. Rev. D \textbf{60}, 109902 (1999)]
%doi:10.1103/PhysRevD.60.109902
[arXiv:gr-qc/9710123 [gr-qc]].
%126 citations counted in INSPIRE as of 16 Feb 2025

%\cite{Kobialko:2022ozq}
\bibitem{Kobialko:2022ozq}
K.~Kobialko, I.~Bogush and D.~Gal'tsov,
``Slice-reducible conformal Killing tensors, photon surfaces, and shadows,''
Phys. Rev. D \textbf{106}, no.2, 024006 (2022)
%doi:10.1103/PhysRevD.106.024006
[arXiv:2202.09126 [gr-qc]].
%14 citations counted in INSPIRE as of 16 Feb 2025

%\cite{Kobialko:2021aqg}
\bibitem{Kobialko:2021aqg}
K.~Kobialko, I.~Bogush and D.~Gal'tsov,
``Killing tensors and photon surfaces in foliated spacetimes,''
Phys. Rev. D \textbf{104} (2021) no.4, 044009.
%doi:10.1103/PhysRevD.104.04400

\bibitem{goldberg1962theorem}
%\cite{goldberg1962theorem}
J.~Goldberg, R.~Sachs, et al., 
``A theorem on Petrov types,'' 
Acta Phys. Pol. B, Proc. Suppl. \textbf{22}, 13 (1962). 

%\cite{Landau:1988}
\bibitem{Landau:1988}
L.~D.~Landau, E.~M.~Lifshitz,
``Textbook on Theoretical Physica. Vol. 2: Classical Field Theory.''

%\cite{Kobialko:2024rqr}
\bibitem{Kobialko:2024rqr}
K.~Kobialko, I.~Bogush and D.~Gal'tsov,
``Uniqueness of the static vacuum asymptotically flat spacetimes with massive particle spheres,''
Phys. Rev. D \textbf{110}, no.4, 044059 (2024)
%doi:10.1103/PhysRevD.110.044059
[arXiv:2406.15127 [gr-qc]].

%\cite{Houri:2007xz,Houri:2008ng,Krtous:2008tb,Frolov:2017kze}
\bibitem{Houri:2007xz}
T.~Houri, T.~Oota and Y.~Yasui,
``Closed conformal Killing-Yano tensor and Kerr-NUT-de Sitter spacetime uniqueness,''
Phys. Lett. B \textbf{656}, 214-216 (2007)
%doi:10.1016/j.physletb.2007.09.034
[arXiv:0708.1368 [hep-th]].
%78 citations counted in INSPIRE as of 24 Jan 2025

%\cite{Houri:2008ng}
\bibitem{Houri:2008ng}
T.~Houri, T.~Oota and Y.~Yasui,``Closed conformal Killing-Yano tensor and uniqueness of generalized Kerr-NUT-de Sitter spacetime,''
Class. Quant. Grav. \textbf{26}, 045015 (2009)
%doi:10.1088/0264-9381/26/4/045015
[arXiv:0805.3877 [hep-th]].
%53 citations counted in INSPIRE as of 24 Jan 2025
%\cite{Krtous:2008tb}

\bibitem{Krtous:2008tb}
P.~Krtous, V.~P.~Frolov and D.~Kubiznak,
``Hidden Symmetries of Higher Dimensional Black Holes and Uniqueness of the Kerr-NUT-(A)dS spacetime,''
Phys. Rev. D \textbf{78}, 064022 (2008)
%doi:10.1103/PhysRevD.78.064022
[arXiv:0804.4705 [hep-th]].
%99 citations counted in INSPIRE as of 24 Jan 2025

%\cite{Aryal:1986sz,Galtsov:1989ct,Hackmann:2010ir,Appels:2017xoe}
\bibitem{Aryal:1986sz}
M.~Aryal, L.~H.~Ford and A.~Vilenkin,
``Cosmic Strings and Black Holes,''
Phys. Rev. D \textbf{34}, 2263 (1986).
%doi:10.1103/PhysRevD.34.2263
%165 citations counted in INSPIRE as of 24 Jan 2025
 
\bibitem{Galtsov:1989ct}
D.~V.~Galtsov and E.~Masar,
``Geodesics in Space-times Containing Cosmic Strings,''
Class. Quant. Grav. \textbf{6}, 1313-1341 (1989).
%doi:10.1088/0264-9381/6/10/004
%53 citations counted in INSPIRE as of 24 Jan 2025

%\cite{Hackmann:2010ir}
\bibitem{Hackmann:2010ir}
E.~Hackmann, B.~Hartmann, C.~Lammerzahl and P.~Sirimachan,
``Test particle motion in the space-time of a Kerr black hole pierced by a cosmic string,''
Phys. Rev. D \textbf{82}, 044024 (2010)
%doi:10.1103/PhysRevD.82.044024
[arXiv:1006.1761 [gr-qc]].
%51 citations counted in INSPIRE as of 24 Jan 2025

%\cite{Appels:2017xoe}
\bibitem{Appels:2017xoe}
M.~Appels, R.~Gregory and D.~Kubiznak,
``Black Hole Thermodynamics with Conical Defects,''
JHEP \textbf{05}, 116 (2017)
%doi:10.1007/JHEP05(2017)116
[arXiv:1702.00490 [hep-th]].
%89 citations counted in INSPIRE as of 24 Jan 2025
\bibitem{Clement:2015aka}
G.~Cl\'ement, D.~Gal'tsov and M.~Guenouche,
``NUT wormholes,''
Phys. Rev. D \textbf{93}, no.2, 024048 (2016)
%doi:10.1103/PhysRevD.93.024048
[arXiv:1509.07854 [hep-th]].
%73 citations counted in INSPIRE as of 02 Mar 2025

%\cite{Clement:2022pjr}
\bibitem{Clement:2022pjr}
G.~Cl\'ement and D.~Gal'tsov,
``Rotating traversable wormholes in Einstein-Maxwell theory,''
Phys. Lett. B \textbf{838}, 137677 (2023)
%doi:10.1016/j.physletb.2023.137677
[arXiv:2210.08913 [gr-qc]].
%15 citations counted in INSPIRE as of 02 Mar 2025

%\cite{Corral:2024lva}
\bibitem{Corral:2024lva}
C.~Corral and R.~Olea,
``Electric-magnetic duality of dyonic Kerr-Newman-NUT-AdS spacetimes,''
Phys. Rev. D \textbf{110}, no.10, 104021 (2024)
%doi:10.1103/PhysRevD.110.104021
[arXiv:2408.03901 [hep-th]].
%0 citations counted in INSPIRE as of 26 Feb 2025


%\cite{Caldarelli:1999xj}
\bibitem{Caldarelli:1999xj}
M.~M.~Caldarelli, G.~Cognola and D.~Klemm,
``Thermodynamics of Kerr-Newman-AdS black holes and conformal field theories,''
Class. Quant. Grav. \textbf{17}, 399-420 (2000)
%doi:10.1088/0264-9381/17/2/310
[arXiv:hep-th/9908022 [hep-th]].
%842 citations counted in INSPIRE as of 14 Feb 2025

%\cite{Aliev:2007qi}
\bibitem{Aliev:2007qi}
A.~N.~Aliev,
``Electromagnetic Properties of Kerr-Anti-de Sitter Black Holes,''
Phys. Rev. D \textbf{75}, 084041 (2007)
%doi:10.1103/PhysRevD.75.084041
[arXiv:hep-th/0702129 [hep-th]].
%76 citations counted in INSPIRE as of 14 Feb 2025
%%%%%%%%%%%%%%%%%%%%%%%%%%%%%%%%%%%%%%%%%%%%%%%%%%%%%%%%%%%%%%%%%%%%%%%%%%%%%%%%%%%%%%%%%%%%%%%%%%%%%%%%%%%%%%%%%%%%
\end{thebibliography}
\end{document}